# A Compact Mathematical Model of the World System Economic and Demographic Growth, 1 CE – 1973 CE

***Korotayev, Andrey V.***, Russian State University for the Humanities, Faculty of Global Studies of the Moscow State University, Oriental Institute and Institute for African Studies, Russian Academy of Sciences, Moscow, and "Complex System Analysis and Mathematical Modeling of the World Dynamics" Project, Russian Academy of Sciences
***Malkov, Artemy S.***, Institute of Economics of the Russian Academy of Sciences

*Abstract*

*We propose an extremely simple mathematical model that is shown to be able to account for more than 99 per cent of all the variation in economic and demographic macrodynamics of the world for almost two millennia of its history. This appears to suggest a novel approach to the formation of the general theory of social macroevolution.*

## Introduction

In 1960 von Foerster, Mora, and Amiot conducted a statistical analysis of the available world population data and found out that the general shape of the world population ($N$) growth is best approximated by the curve described by the following equation:

$$N = \frac{C}{t_0 - t}, \qquad (1)$$

where $C$ and $t_0$ are constants, whereas $t_0$ corresponds to an absolute limit of such a trend at which $N$ would become infinite, and thus logically implies the certainty of the empirical conclusion that further increases in the growth trend will cease well before that date, which von Foerster wryly called the “doomsday” implication of power-law growth – he refers tongue-in-cheek to the estimated $t_0$ as "Doomsday, Friday, 13 November, A.D. 2026" [1]. (Of course, von Foerster and his colleagues did not imply that the world population on that day could actually become infinite. The real implication was that the world population growth pattern that was followed for many centuries prior to 1960 was about to come to an end and be transformed into a radically different pattern. Note that this prediction began to be fulfilled only in a few years after the "Doomsday" paper was published, since the 1960 the world population growth began to diverge more and more from the blow-up regime, and now it is not hyperbolic any more – see, *e.g.*, Ref. [2], where we present a compact mathematical model that describes both the hyperbolic development of the World System in the period prior to the early 1970s, and its withdrawal from the blow-up regime in the subsequent period).

Von Foerster, Mora, and Amiot try to account for their empirical observations by modifying the usual starting equations (2) and (3) for population dynamics, so as to describe the process under consideration:

$$\frac{dN}{dt} = B - D, \qquad (2)$$

where $N$ is the number of people, $B$ is the number of births, and $D$ is the number of deaths in the unit of time;

$$\frac{dN}{dt} = (a_1 N) - (a_2 N + bN^2), \qquad (3)$$

where $a_1N$ corresponds to the number of births $B$, and $a_2N + bN^2$ corresponds to the number of deaths in equation (2); $r$, $K$, $a_1$, $a_2$, $b$ are positive coefficients connected between themselves by the following relationships:

$$r = a_1 - a_2 \quad \text{and} \quad b = \frac{r}{K}, \qquad (4)$$

They start with the observations that when individuals in a population compete in a limited environment, the growth rate typically *decreases* with the greater number $N$ in competition. This situation would typically apply where sufficient communication is lacking to enable resort to other than a competitive and nearly zero-sum multiperson game. It might not apply, they suppose, when the elements in a population “possess a system of communication which enables them to form coalitions” and especial-ly when “all elements are so strongly linked that the population as a whole can be considered from a game-theoretical point of view as a single person playing a two-person game with nature as the opponent” [1]. Thus, the larger the population ($N^k$ coalition members, where $k \leq 1$) the more the decrease of natural risks and the higher the population growth rate. They suggest modeling such a situation through the introduction of the nonlinearity in the following form:

$$\frac{dN}{dt} = (a_0 N^{\frac{1}{k}})N \,, \tag{5}$$

where $a_0$ and $k$ are constants, which should be determined experimentally. The analysis of experimental data by von Foerster, Mora, and Amiot determines values $a_0 = 5.5\ 10^{-12}$ and $k = 0.99$ that produce the hyperbolic equation for the world population growth:

$$N = N_1 \left( \frac{t_0 - t_1}{t_0 - t} \right)^k , \tag{6}$$

which, assuming $k = 1.0$ [3] is written more shortly as (1) and in equivalent form [4] as (7):

$$\frac{dN}{dt} = \frac{N^2}{C} \,. \tag{7}$$

Note that if von Foerster, Mora, and Amiot had had at their disposal, in addition to the world population data, also the data on the world GDP dynamics for 1–1973 (published, however, only in 2001 by Maddison [5]) they could have made another striking "prediction" – that on Saturday, 23 July, 2005 an "economic doomsday" would take place, that is on that day the world GDP would become infinite. They would have also found that in 1–1973 CE the world GDP growth had followed quadratic-hyperbolic rather than simple hyperbolic pattern.

Indeed, Maddison's estimates of the world GDP dynamics for 1–1973 CE are almost perfectly approximated by the following equation:

$$G = \frac{\mathrm{C}}{(t_0 - t)^2} \,, \tag{8}$$

where $G$ is the world GDP, $C = 17355487.3$ and $t_0 = 2005.56$ (see Fig. 1):

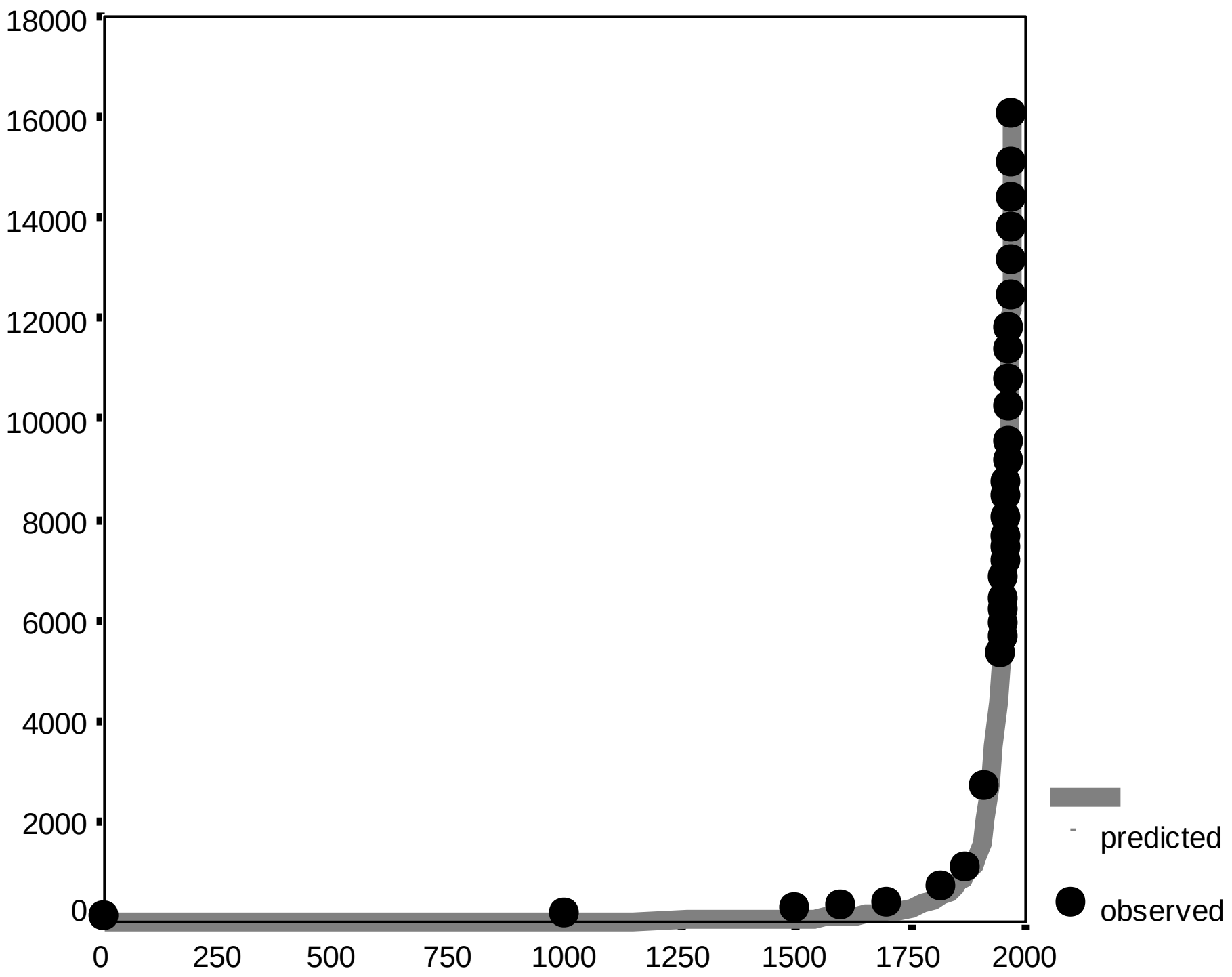

Fig. 1. World GDP dynamics, 1–1973 CE (in billions of 1990 international dollars, PPP): the fit between predictions of quadratic-hyperbolic model and the observed data. NOTE: $R$ = .9993, $R^2$ = .9986, $p$ << .0001. The black markers correspond to Maddison's [5] estimates (Maddison's estimates of the world per capita GDP for 1000 CE has been corrected on the basis of Meliantsev [6]). The grey solid line has been generated by the following equation:

$G = \frac{17749573.1}{(2006-t)^2}$. The best-fit values of parameters $C$ (17749573.1) and $t_0$ (2006) have been calculated with the least squares method (actually, as was mentioned above, the best fit is achieved with $C$ = 17355487.3 and $t_0$ = 2005.56 [which gives just the "doomsday Saturday, 23 July, 2005"], but we have decided to keep hereafter to the integer year numbers).

In fact, the fit provided by the simple hyperbolic model for the world GDP dynamics for 1–1973 CE is also in no way bad, but still less perfect than the one found for the quadratic hyperbolic model (see Fig. 2):

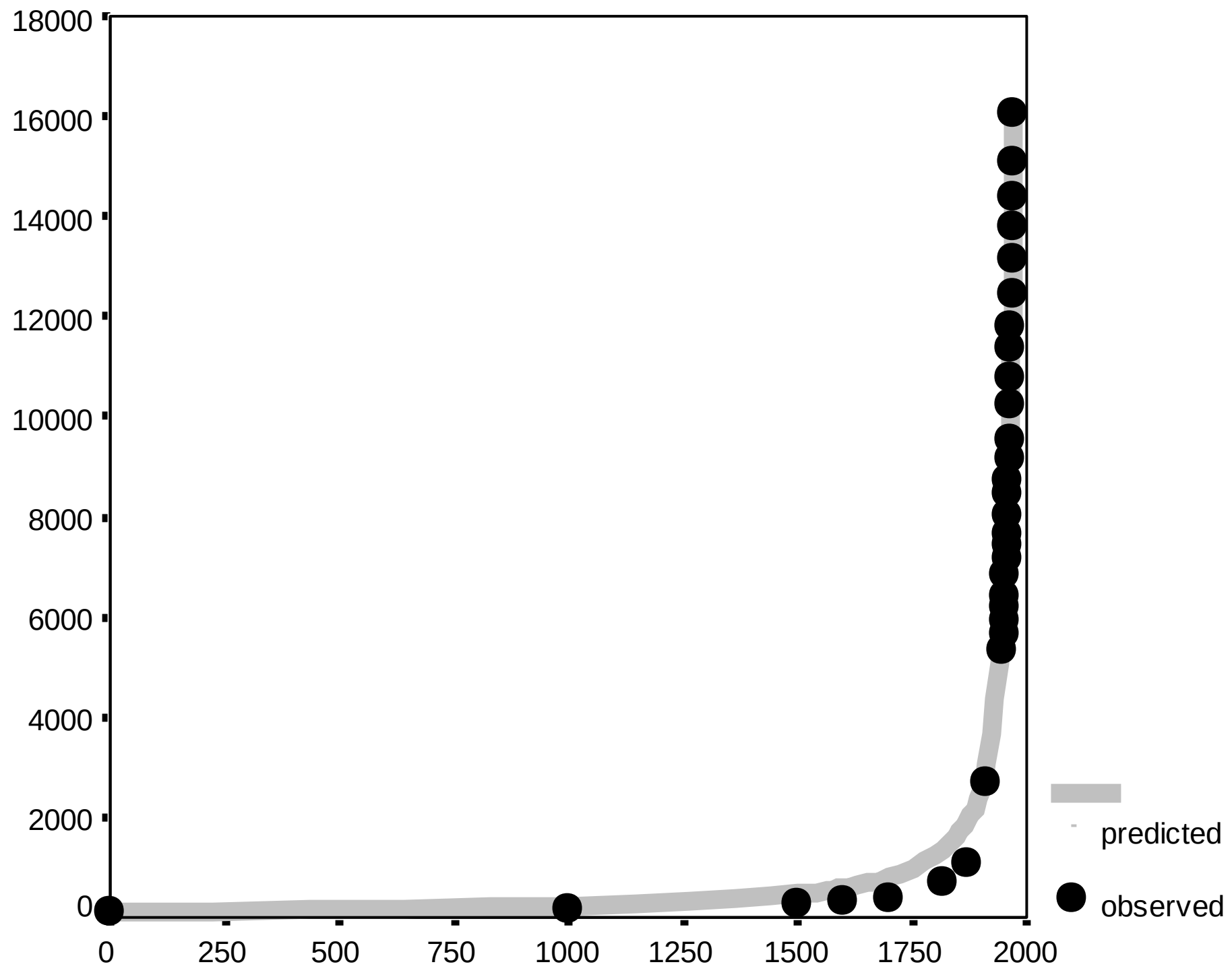

Fig. 2. World GDP dynamics, 1–1973 CE (in billions of 1990 international dollars, PPP): the fit between predictions of simple hyperbolic model and the observed data. NOTE: $R$ = .9978, $R^2$ = .9956, $p$ << .0001. The black markers correspond to Maddison's [5] estimates (Maddison's estimates of the world per capita GDP for 1000 CE has been corrected on the basis of Meliantsev [6]). The grey solid line has been generated by the following equation:

$G = \frac{227906.1}{1987 - t}$. The best-fit values of parameters $C$ (227906.1) and $t_0$ (1987) have been calculated with the least squares method.

This, of course, suggests that the long-term dynamics of the world GDP up to the 1970s should be approximated better with a quadratic rather than simple hyperbola.

Note that, on the other hand, the simple hyperbolic model does fit the world population dynamics in 1–1973 much better than the world GDP one. Indeed, though the world 1–1973 CE GDP dynamics fits the simple hyperbolic model of type (1) rather well ($R$ = .9978, $R^2$ = .9956, $p$ << .0001), this fit is still significantly worse than the one observed for the world population growth for the same time ($R$ = .9996, $R^2$ = .9991, $p$ << .0001) (see Fig. 3):

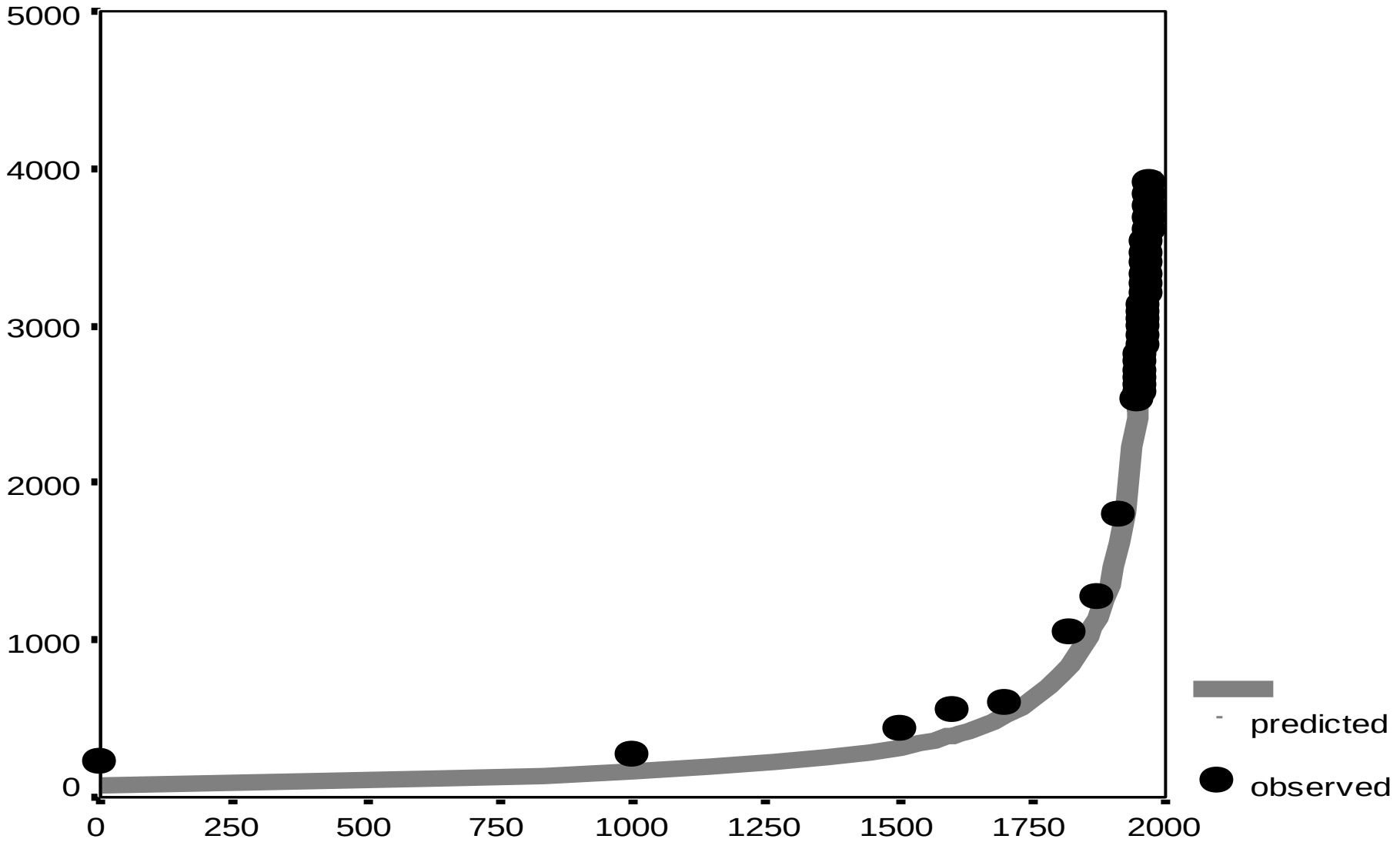

Fig. 3. World population dynamics, 1–1973 CE (in millions): the fit between predictions of hyperbolic model and the observed data. NOTE: $R$ = .9996, $R^2$ = .9991, $p$ << .0001. The black markers correspond to Maddison's [5] estimates. The grey solid line has been generated by the following equation: $N = \frac{163158.78}{2014 - t}$. The best-fit values of parameters $C$ (163158.78) and $t_0$ (2014) have been calculated with the least squares method.

However, the quadratic hyperbolic model renders a worse fit for the world population ($R$ = .9982, $R^2$ = .9963, $p$ << .0001) (see Fig. 4):

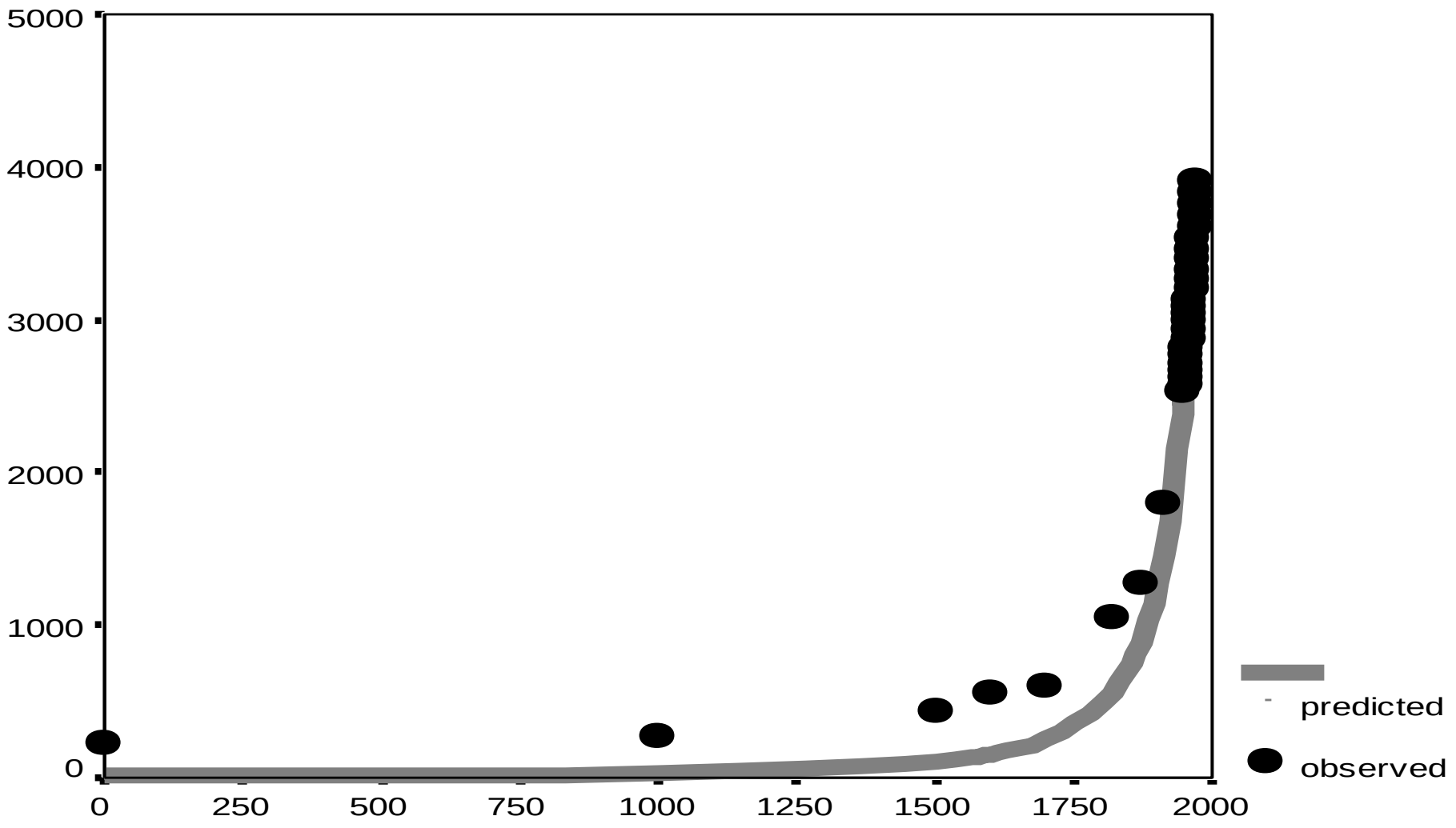

Fig. 4. World population dynamics, 1–1973 CE (in millions): the fit between predictions of quadratic-hyperbolic model and the observed data. NOTE: $R$ = .9982, $R^2$ = .9963, $p$ << .0001. The black markers

correspond to Maddison's [5] estimates. The grey solid line has been generated by the following equation:
$N = \frac{33505220.7}{(2065-t)^2}$. The best-fit values of parameters $C$ (33505220.702) and $t_0$ (2065) have been calculated with the least squares method.

In this article we show that it is in no way coincidental that the world GDP dynamics in 1–1973 is approximated so well with a quadratic hyperbola, whereas the world population one does with a simple hyperbola. We also suggest a compact explanatory mathematical model describing the world population and GDP growth in 1–1973 CE and discuss its implications.

We believe that the most significant progress towards the development of a compact mathematical model describing the world GDP growth has been achieved by Michael Kremer [7], whose model will be discussed next (for the other mathematical models describing the world hyperbolic growth see [3,4,8,9,10,11,12,13,14,15]; of special interest are the mathematical models developed by Anders Johansen and Didier Sornette [16]; see also [17]; the explanatory logic used by Michael Kremer appears to be first suggested by Rein Taagapera [18,19]).

## Michael Kremer's model of the world demographic and technological growth

One of the basic assumptions of Kremer's model was suggested already in the 18th century by Thomas Malthus [20]. It may be worded in the following way:

"Population is limited by the available technology, so that the growth rate of population is proportional to the growth rate of technology" [7]. (8)

The basic model suggested by Kremer assumes that the production output depends on just two factors: technological level and population size. Kremer uses the following symbols to denote respective variables: $Y$ – output, $p$ – population, $A$ – the level of technology, etc.; while describing Kremer's models we will employ symbols (closer to Kapitza's ones) used in our model, naturally, without distorting the sense of Kremer's equations.

Kremer assumes that overall output produced by the world economy equals

$$G = TN^{\alpha}V^{1-\alpha}$$

where $G$ is output, $T$ is the level of technology, $V$ is land, $0 < \alpha < 1$ is a parameter. Actually Kremer uses a variant of the Cobb-Douglas production function. Kremer further qualifies that variable $V$ is normalized to one. The resultant equation for output looks as follows:

$$G = rTN^{\alpha}, \tag{9}$$

where $r$ and $\alpha$ are constants.

Further Kremer uses statement (8), formulating it in the following way: "In this simplified model I assume that population adjusts instantaneously to $\overline{N}$" [7]. Value $\overline{N}$ in this model corresponds to population size, at which it produces equilibrium level of per capita income $\overline{g}$, whereas "population increases above some steady state equilibrium level of per capita income, $\overline{g}$, and decreases below it" [7].

Thus, equilibrium level of population $\overline{N}$ equals

$$\overline{N} = \left(\frac{\overline{g}}{T}\right)^{\frac{1}{\alpha-1}}. \tag{10}$$

Hence, the equation for population size is not actually dynamic. In Kremer's model dynamics is put into the equation for technological growth. Kremer uses the following assumption of the Endogenous Technological Growth theory [21,22,23,24,25,26]:

"High population spurs technological change because it increases the number of potential inventors…[1]. All else equal, each person's chance of inventing something is independent of population. Thus, in a larger population there will be proportionally more people lucky or smart enough to come up with new ideas" [7]; thus, "the growth rate of technology is proportional to total population" [7]. (11)

As this supposition, up to our knowledge, was first proposed by Simon Kuznets [21], we shall denote the respective type of dynamics as "Kuznetsian", whereas the systems where the "Kuznetsian" population-technological dynamics is combined with "Malthusian" demographic one will be denoted as "Malthusian-Kuznetsian".

This assumption is expressed mathematically by Kremer in the following way:

[1] "This implication flows naturally from the nonrivalry of technology… The cost of inventing a new technology is independent of the number of people who use it. Thus, holding constant the share of resources devoted to research, an increase in population leads to an increase in technological change" [7].

$$\frac{dT}{dt} : T = bN \, , \qquad (12)$$

where $b$ is average innovating productivity per person.

Note that this implies that the dynamics of absolute technological growth rate can be described by the following equation:

$$\frac{dT}{dt} = bNT \, . \qquad (13)$$

Kremer further combines the research and population determination equations in the following way:

"Since population is limited by technology, the growth rate of population is proportional to the growth rate of technology. Since the growth rate of technology is proportional to the level of population, the growth rate of population must also be proportional to the level of population. To see this formally, take the logarithm of the population determination equation, [(10)], and differentiate with respect to time:

$$\frac{dN}{dt} : N = \frac{1}{1-\alpha} \left( \frac{dT}{dt} : T \right) .$$

Substitute in the expression for the growth rate of technology from [(12)], to obtain

$$\frac{dN}{dt} : N = \frac{g}{1-\alpha} N \text{ " [7].} \qquad (14)$$

Note that multiplying both parts of equation (14) by $N$ we get

$$\frac{dN}{dt} = aN^2 \, , \qquad (7')$$

where $a$ equals

$$a = \frac{g}{1-\alpha} \, .$$

Of course, the same equation can be also written as

$$\frac{dN}{dt} = \frac{N^2}{C} \, , \qquad (7)$$

where $C$ equals

$$C = \frac{1-\alpha}{g}.$$

Thus, Kremer's model produces precisely the same dynamics as the ones of von Foerster and Kapitza (and, consequently, it has just the same phenomenal fit with the observed data). However, it also provides a very convincing explanation WHY throughout most of the human history the absolute world population growth rate tended to be proportional to $N^2$. Within both models the growth of population from, say, 10 million to 100 million will result in the growth of *dN/dt* 100 times. However, von Foerster and Kapitza failed to explain convincingly why *dN/dt* tended to be proportional to $N^2$. Kremer's model explains this in what seems to us a rather convincing way (though Kremer himself does not appear to have spelled this out in a sufficiently clear way). The point is that the growth of the world population from 10 to 100 million implies that the human technology also grew approximately 10 times (as it turns out to be able to support a ten times larger population). On the other hand, the growth of population 10 times also implies 10-fold growth of the number of potential inventors, and, hence, 10-fold increase in the relative technological growth rate. Hence, the absolute technological growth will grow 10 x 10 = 100 times (in accordance to equation (13)). And as *N* tends to the technologically determined carrying capacity ceiling, we have all grounds to expect that *dN/dt* will also grow just 100 times.

## World dynamics as the World System dynamics

Note that Kremer's model suggests ways to answer one of the main objections raised against the models of the world population hyperbolic growth. Indeed, by the moment the mathematical models of the world population hyperbolic growth have not been accepted by the social science academic community (The title of an article by a social scientist discussing Kapitza's model, "Demographic Adventures of a Physicist" [27], is rather telling in this respect). We believe there are substantial reasons for such a position, and the authors of the respective models are to blame for their rejection to no less extent than social scientists.

Indeed, all the respective models are based on an assumption that the world population can be treated as an integrated system for many centuries, if not millennia before 1492 (Actually, von Foerster, Mora, and Amiot [1] detected the hyperbolic growth pattern for 1–1958 CE; however, Kremer [7] suggests that it can be traced up to 1 million BCE, whereas Kapitza [4] insists that this pattern is even much more ancient). Already, von Foerster, Mora, and Amiot spelled out this assumption in a rather explicit way:

"However, what may be true for elements which, because of lack of adequate communication among each other, have to resort to a competitive, (almost) zero-sum multiperson game may be false for elements that possess a system of communication which enables them to form coalitions until all elements are so strongly linked that the population as a whole can be considered from a game-theoretical point of view as a single person playing a two-person game with nature as its opponent" [1].

However, did, e.g. in 1–1500 CE, the inhabitants of, say, Central Asia, Tasmania, Hawaii, Terra del Fuego, Kalahari etc. (that is, just the world population) really have such an "adequate communication", which made "all elements… so strongly linked that the population as a whole can be considered from a game-theoretical point of view as a single person playing a two-person game with nature as its opponent"? For any historically minded social scientist the answer to this question is perfectly clear. And, of course, this answer is squarely negative. Against this background it is hardly surprising that those social scientists who managed to get across the world population hyperbolic growth models had sufficient grounds to treat them just as "demographic adventures of physicists" (note, that indeed seven out of ten currently known authors of such models are just physicists), as none of the respective authors [1,3,4,7,8,9,10,16] has provided any answer to the question above.

However, it is not so difficult to provide such an answer.

The hyperbolic trend observed for the world population growth after 10000 BCE appears to be mostly a product of the growth of the World System, which seems to have originated in the West Asia around that time in direct connection with the Neolithic Revolution.[2] The presence of the hyperbolic trend indicates that the major part of the entity in question had some systemic unity, and the evidence for this unity is readily available. Indeed, we have evidence for the systematic spread of major innovations (domesticated cereals, cattle, sheep, goats, horses, plow, wheel, copper, bronze, and later iron technology, and so on) throughout the whole North African – Eurasian Oikumene for a few millennia BCE (see, e.g., Ref. [28] for a synthesis of such evidence). As a result the evolution of societies of this part of the world already at this time cannot be regarded as truly independent. By the end of the 1st millennium BCE we observe a belt of cultures stretching from the Atlantic to the Pacific with an astonishingly similar level of cultural complexity based on agriculture involving production of wheat and other specific cereals, cattle, sheep, goats, based on plow, iron metallurgy, wheeled transport, professional armies with rather similar weapons, cavalries, developed bureaucracies and Axial Age ideologies and so on – this list can be extended for pages). A few millennia before we would find a belt of societies with a similarly strikingly close level and character of cultural complexity stretching from the Balkans up to the Indus Valley outskirts – note that in both cases the respective entities included the major part of the contemporary world population (see, e.g. [29,30]). We would interpret this as a tangible result of the World System functioning. The alternative explanations would involve a sort of miraculous scenario – that the cultures with a strikingly similar levels and character of complexity somehow developed independently from each other in a very large but continuous zone, whereas nothing like that appeared in the other parts of the world, which were not parts of the World System. We find such an alternative explanation highly implausible.

[2] We are inclined to speak together with Frank (e.g., [31]), but not with Wallerstein [32] about the single World System, which originated long before the "long 16th century".

Thus, we would tend to treat the world population hyperbolic growth pattern as reflecting the growth of quite a real entity, the World System.

A few other points seem to be relevant here. Of course, there would be no grounds to speak about the World System stretching from the Atlantic to the Pacific even at the beginning of the 1$^{st}$ Millennium CE if we applied the "bulk-good" criterion suggested by Wallerstein [32], as there was no movement of bulk goods at all between, say, China and Europe at this time (as we have no grounds not to agree with Wallerstein in his classification of the 1$^{st}$ century Chinese silk reaching Europe as a luxury, rather than bulk good). However, the 1$^{st}$ century CE (and even the 1$^{st}$ millennium BCE) World System would be definitely qualified as such if we apply a "softer" information network criterion suggested by Chase-Dunn and Hall [33]. Note that at our level of analysis the presence of an information network covering the whole of World System is a perfectly sufficient condition, which makes it possible to consider this system as a single evolving entity. Yes, in the 1$^{st}$ millennium BCE any bulk goods could hardly penetrate from the Pacific coast of Eurasia to its Atlantic coast. However, the World System has reached by that time such a level of integration that the iron metallurgy could spread through the whole of the World System within a few centuries.

Yes, in the millennia preceding the European colonization of Tasmania its population dynamics – oscillating around 4000 level (*e.g.*, [28]) were not influenced by the World System population dynamics and did not influence it at all. However, such facts just suggest that since the 10$^{th}$ millennium BCE the dynamics of the world population reflects very closely just the dynamics of the World System population (see Refs. [34,35] for more detail).

## A compact mathematical model of the economic and demographic development of the World System

Though Kremer's model provides a virtual explanation how the World System techno-economic development in connection with the demographic dynamics could lead to the hyperbolic population growth, Kremer did not specify his model to such an extent that it could also describe the economic development of the World System and that such a description could be tested empirically.

In fact, it appears possible to propose a very simple mathematical model describing both the economic and demographic development of the World System up to 1973 using the same assumptions as the ones employed by Kremer.

Kremer's analysis suggests the following relationship between per capita GDP and population growth rate (see Fig. 5):

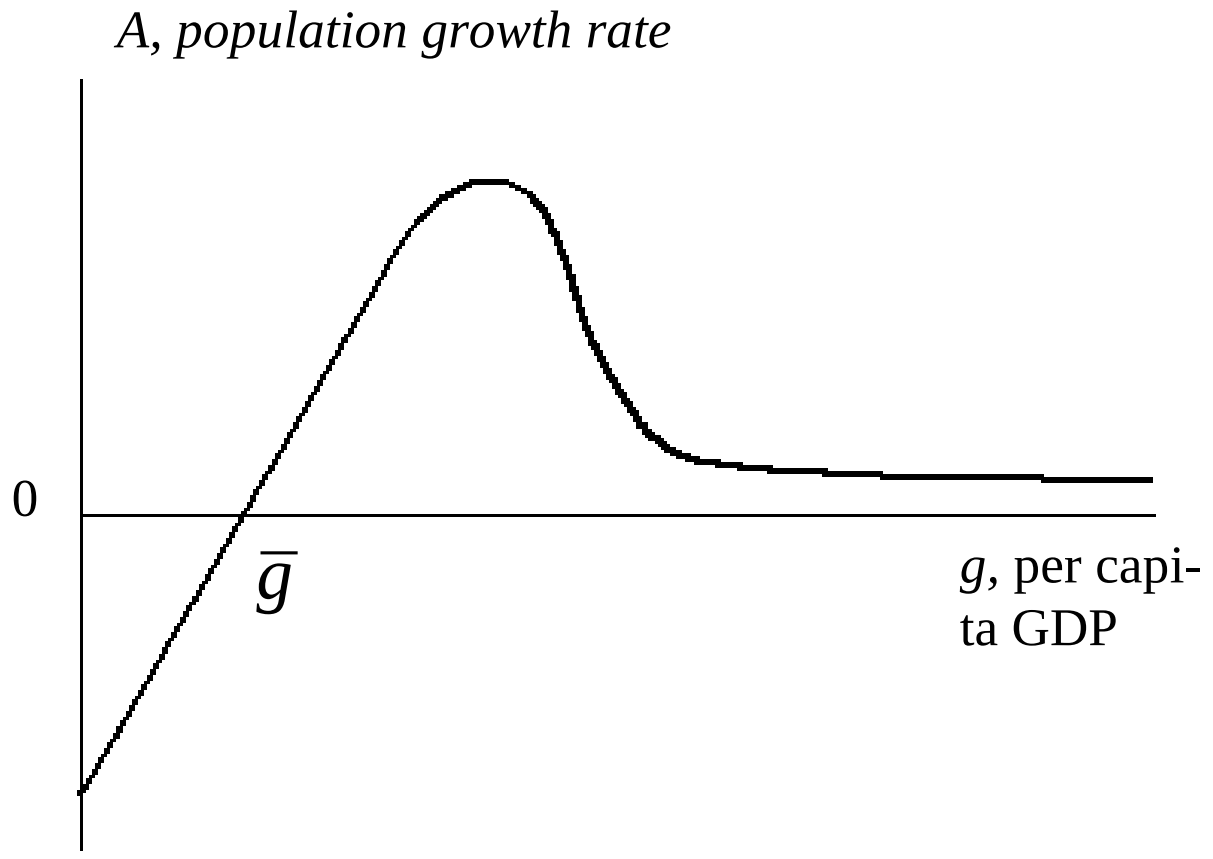

Fig. 5. Relationship between per capita GDP and Population Growth Rate according to Kremer [7]

This suggests that for low per capita GDP range the influence of this variable dynamics on the population growth can be described with the following equation:

$$\frac{dN}{dt} = aSN\ , \qquad (17)$$

where *S* is surplus, which is produced per person over the amount (*m*), which is minimally necessary to reproduce the population with a zero growth rate in a Malthusian system (thus, $S = g - m$, where *g* denotes per capita GDP).

Note, that this model generates some predictions, which could be tested empirically. For example, the model predicts that the relative world population growth ( $r_N = \frac{dN}{dt} : N$ ) should be lineally proportional to the world per capita surplus production:

$$r_N = aS \; . \qquad (18)$$

The empirical test of this hypothesis has supported it, the respective correlation has turned out to be in the predicted direction, very strong ($R$ = .961), and significant beyond any doubt ($p$ = 0.00004) (see Fig. 6):

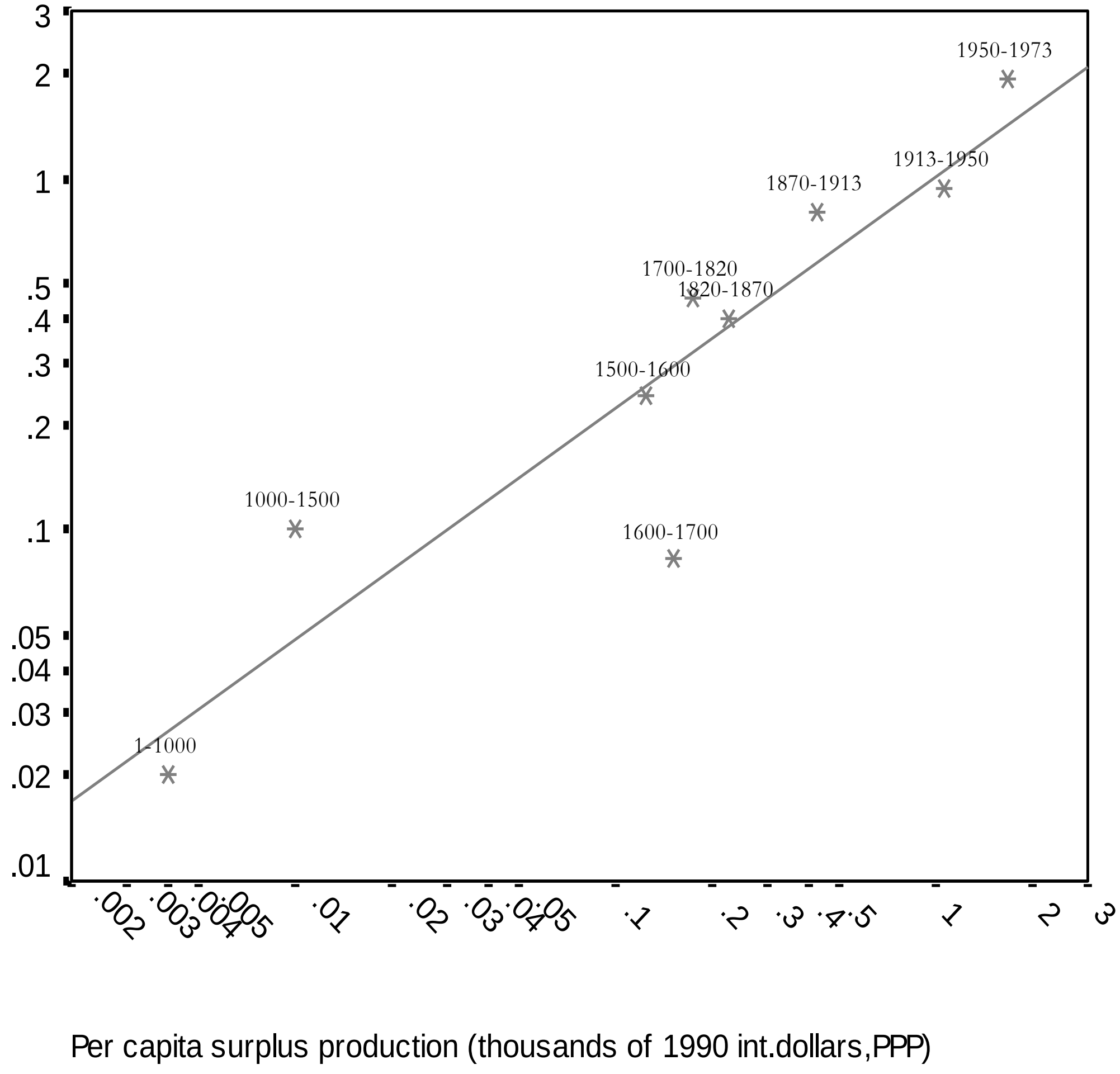

Fig. 6. Correlation between the per capita surplus production and world population growth rates for 1–1973 CE (scatterplot with fitted regression line). NOTES: $R$ = .961, $p$ = 0.00004. Data source – [5]; Maddison's estimates of the world per capita GDP for 1000 CE has been corrected on the basis of Me-

liantsev [6]. *S* values were calculated on the basis of *m* estimated as 440 international 1990 dollars in purchasing power parity (PPP); for the justification of this estimate see Ref. [41].

The mechanisms of this relationship are perfectly evident. In the range \$440–3500[3] the per capita GDP growth leads to very substantial improvements in nutrition, health care, sanitation etc. resulting in a precipitous decline of death rates (see, *e.g.*, Fig. 7):

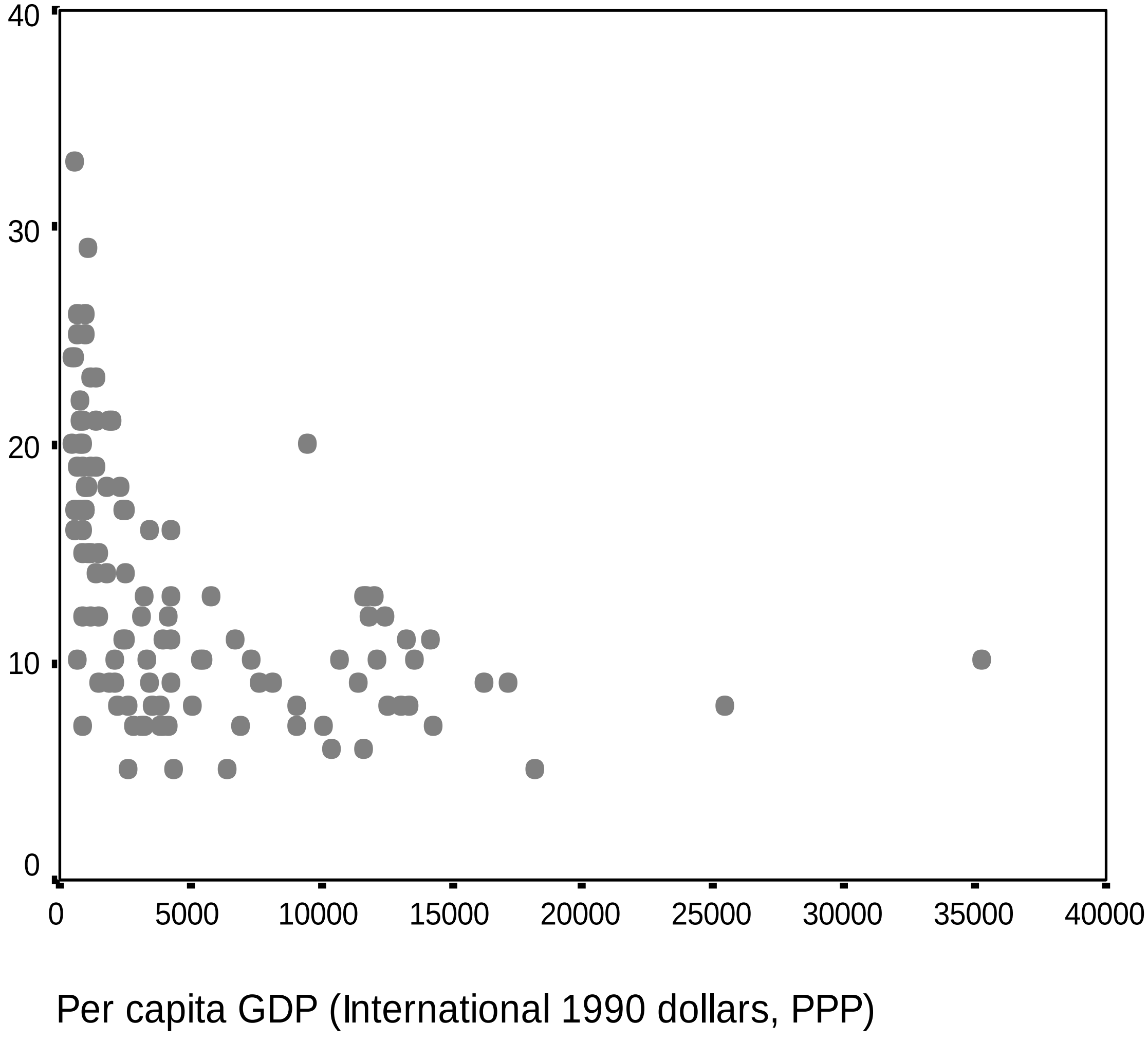

Fig. 7. Correlation between per capita GDP and death rate for countries of the world in 1975NOTE: data sources – [5] (for per capita GDP), [37] (for death rate).

For example, for 1960 the correlation between per capita GDP and death rate for \$440–3500 range reaches – .634 ($p$ = 0.0000000001) (see Fig. 8):

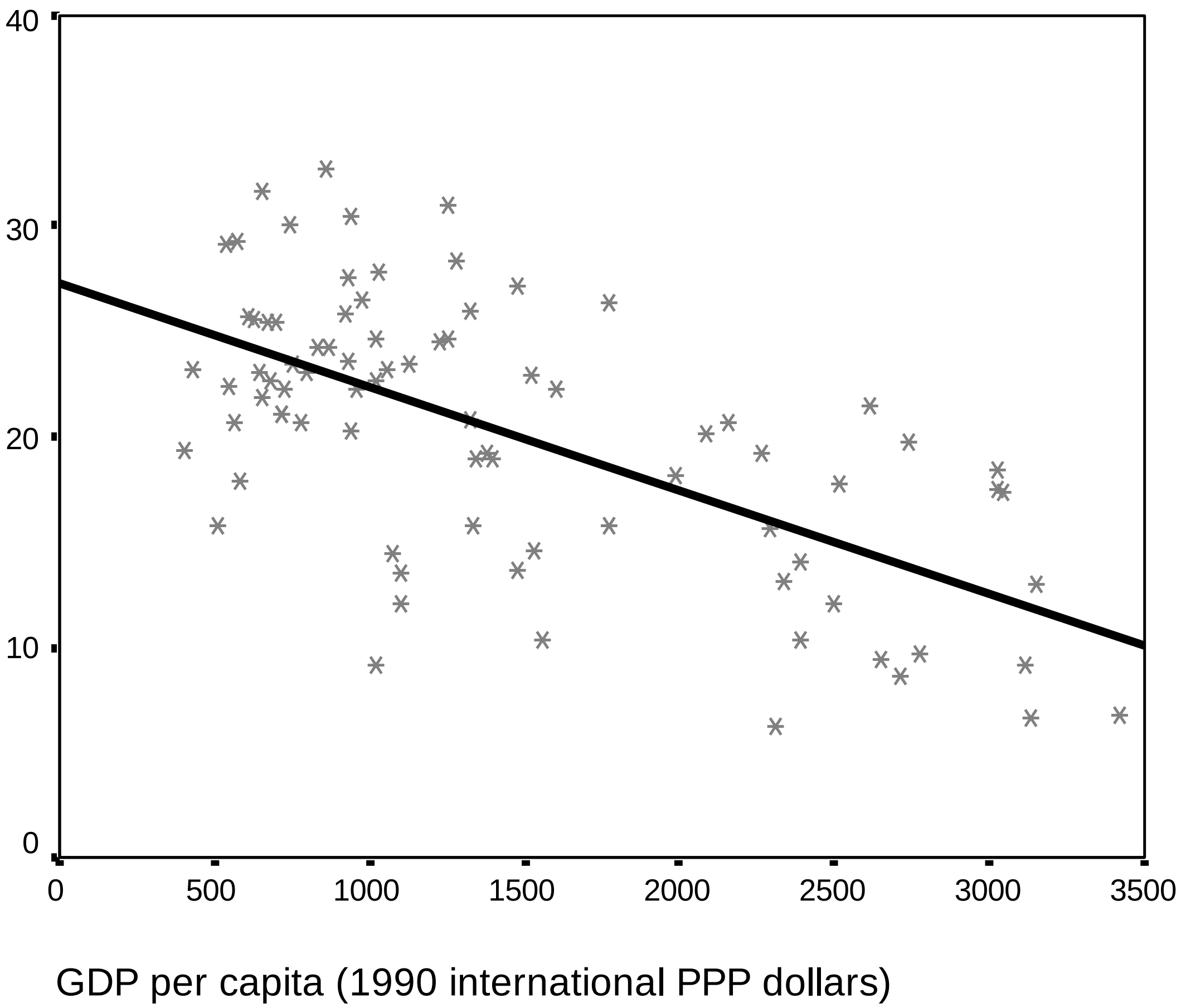

Fig. 8. Correlation between per capita GDP and death rate for countries of the world in 1960 (for $440–3500 range). NOTE: $R = -.634$, $p = 0.0000000001$. Data sources – [5] (for per capita GDP), [37] (for death rate).

Note that during the earliest stages of demographic transition (corresponding just to the range in question) the decline of the death rates is not accompanied by a corresponding decline of the birth rates (*e.g.*, [44]); in fact, they can well even grow (see, *e.g.*, Fig. 9):

[3] Here and throughout the GDP is measured in 1990 international purchasing power parity dollars (after Maddison 2001) if not stated otherwise.

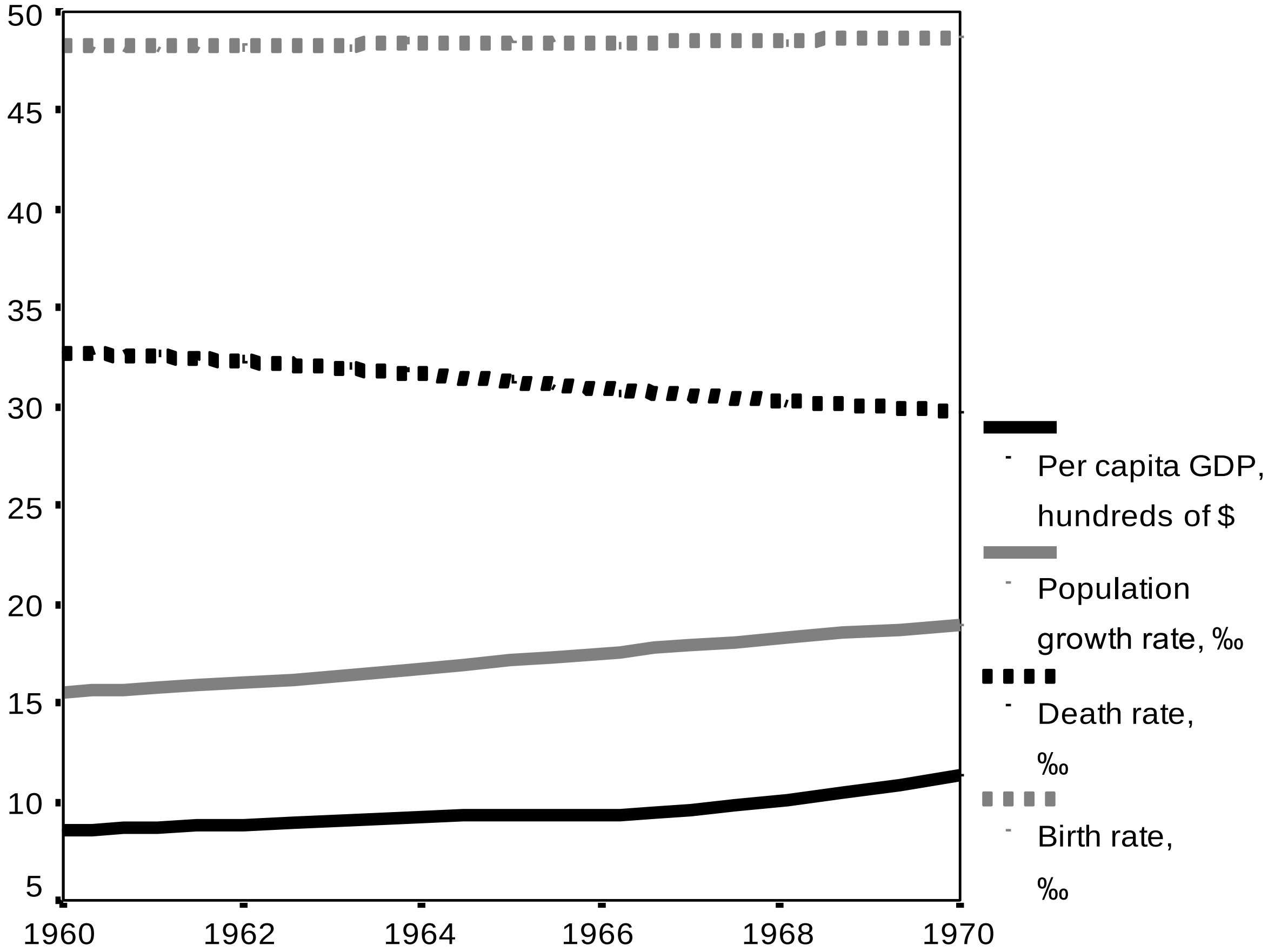

Fig. 9. Economic and demographic dynamics in Sierra Leone, 1960–1970. NOTE: Data sources – [5] for per capita GDP, [37] for demographic variables.

Indeed, say the decline of the death rates from, *e.g.*, 50 to 25‰ implies the growth of life expectancies from 20 to 40 years, whereas a woman within 40 years can bear much more babies than when she dies at the age of 20. In many countries the growth of the fertility rates in the range in question was also connected with the reduction of spacing between births (due to such modernization produced changes as the reduction/abolition of postpartum sex taboos, or the reduction of lactation period connected with the growing availability of foods that can serve as substitutes for maternal milk). Against this background the radical decline of the death rates at the earliest stages of demographic transition is accompanied by as a radical increase of the population growth rates.

Note that equation (17) is not valid for the GDP per capita range > $3500. In this "post-Malthusian" range the death rate reaches bottom levels and then starts slightly growing due to the population aging, whereas the growth of such variables as education (and especially female education), level of social security subsystem development, growing availability of more and more sophisticated family planning

techniques etc. leads to a sharp decline of birth rates (see, *e.g.*, [2,42,43]). As a result in this range the further growth of per capita GDP leads not to the increase of the population growth rates, but to their substantial decrease (see Fig. 10):

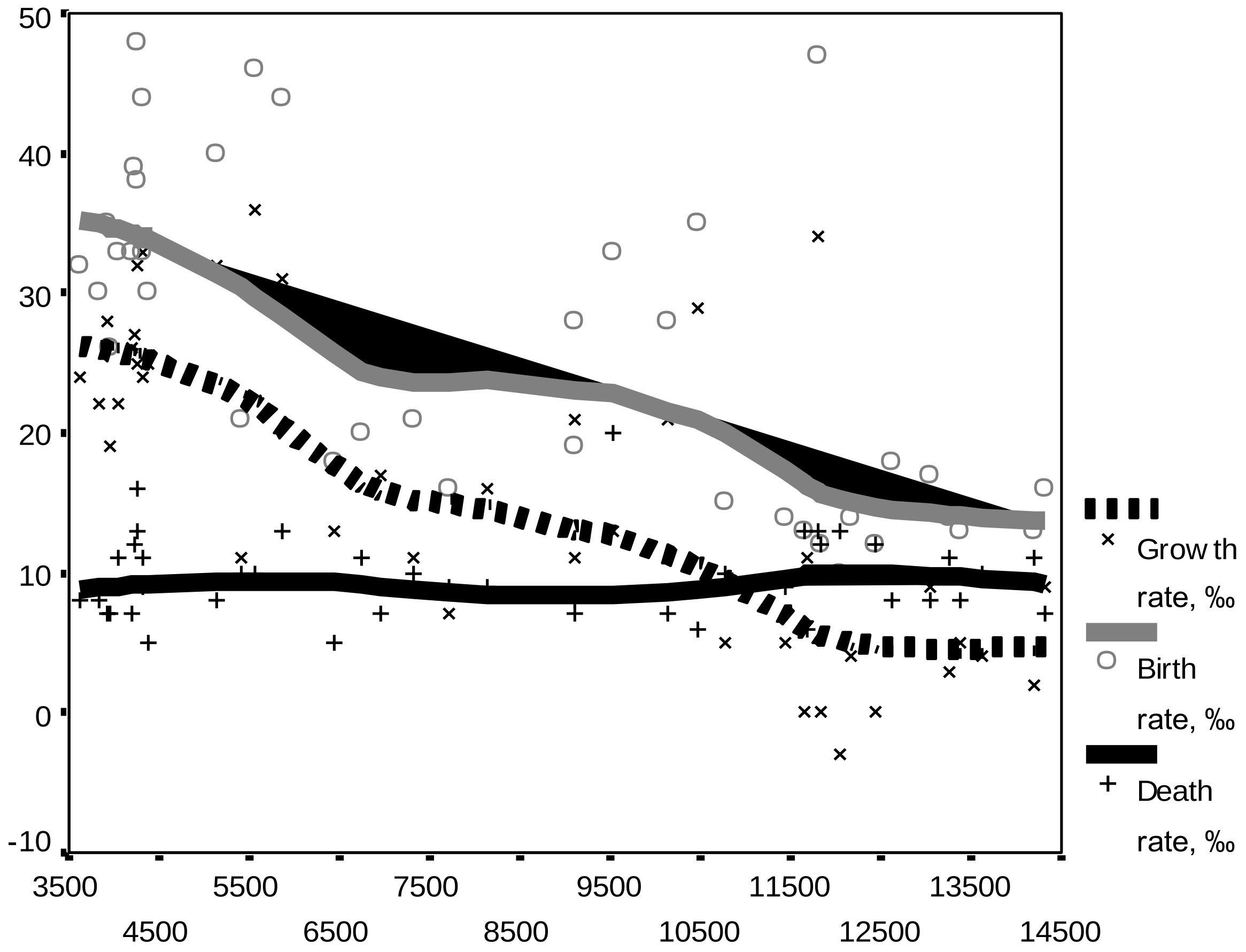

Fig. 10. Relationships between per capita GDP (1990 international PPP dollars, X-axis), Death Rates (‰, Y-axis), Birth Rates (‰, Y-axis), and Population Growth Rates[4] (‰, Y-axis), nations with per capita GDP > $3500, 1975, scatterplot with fitted LOWESS lines. NOTE: Data sources – [5] for per capita GDP), [37] for the other data.

This pattern is even more pronounced for the world countries of 2001, as between 1975 and 2001 the number of countries that had moved out of the first phase of demographic transition to the "post-Malthusian world" substantially increased (see Fig. 11):

[4] Internal ("natural") population growth rate calculated as birth rate minus death rate. We used this variable instead of standard growth rate, as the latter takes into account the influence of emigration and immigration processes, which notwithstanding all their importance are not relevant for the subject of this article, because though they could affect in a most significant way the population growth rates of particular countries, they do not affect the world population growth.

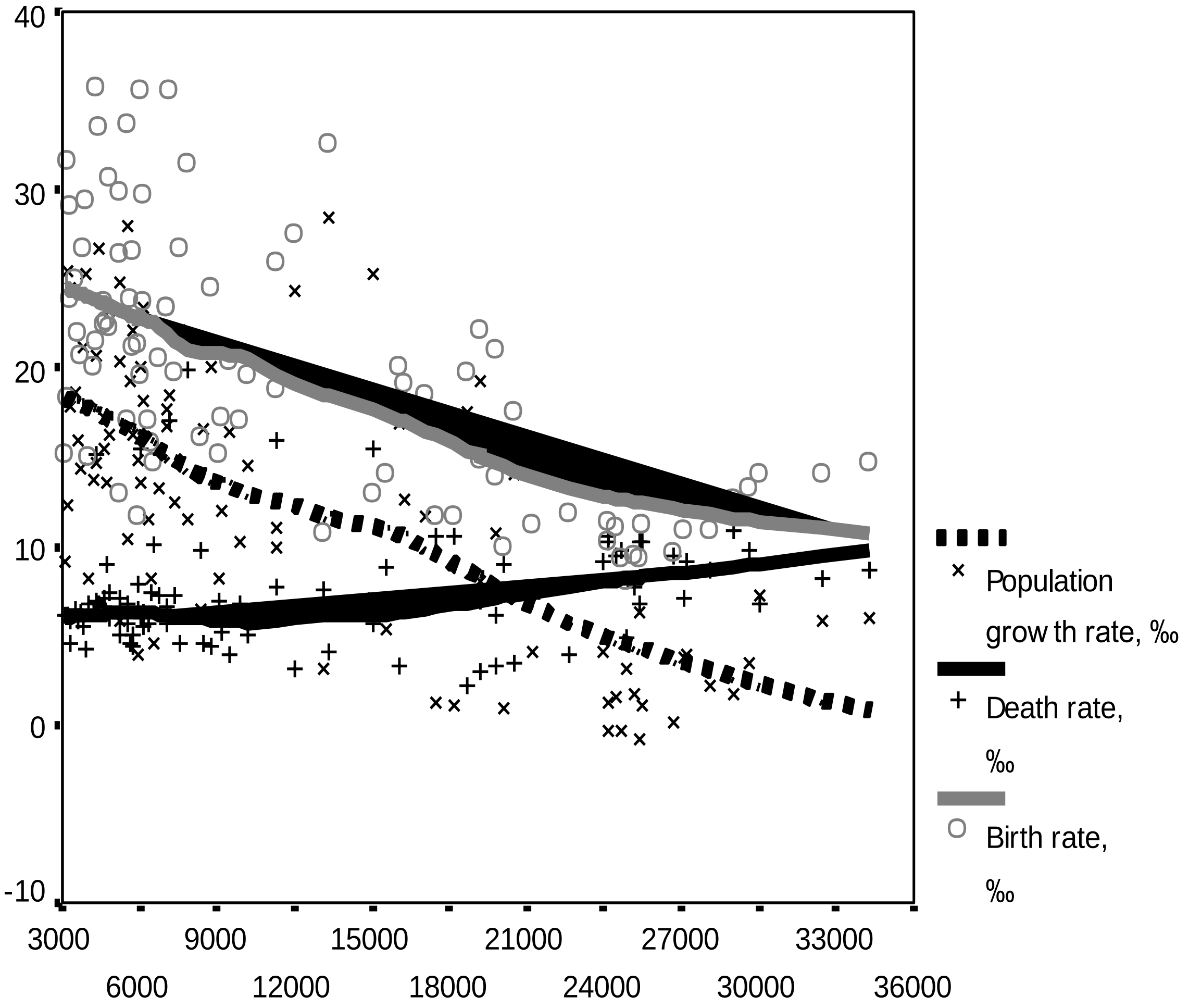

Fig. 11. Relationships between per capita GDP (2001 PPP USD, X-axis), Death Rates (‰, Y-axis), Birth Rates (‰, Y-axis), and Population Growth Rates (‰, Y-axis), nations with per capita GDP > $3000, 2001, scatterplot with fitted LOWESS lines. NOTE: Data source – [37]. Omitting the former Communist countries of Europe, which are characterized by a very specific pattern of demographic growth (or, in fact, to be more exact, demographic decline – see, e.g., [39, 40]).

It is also highly remarkable that as soon as the world per capita GDP approached $3000 and exceeded this (thus, with $S$ ~ $2500), the positive correlation between $S$ and $r$ first dropped to zero (see Fig. 12):

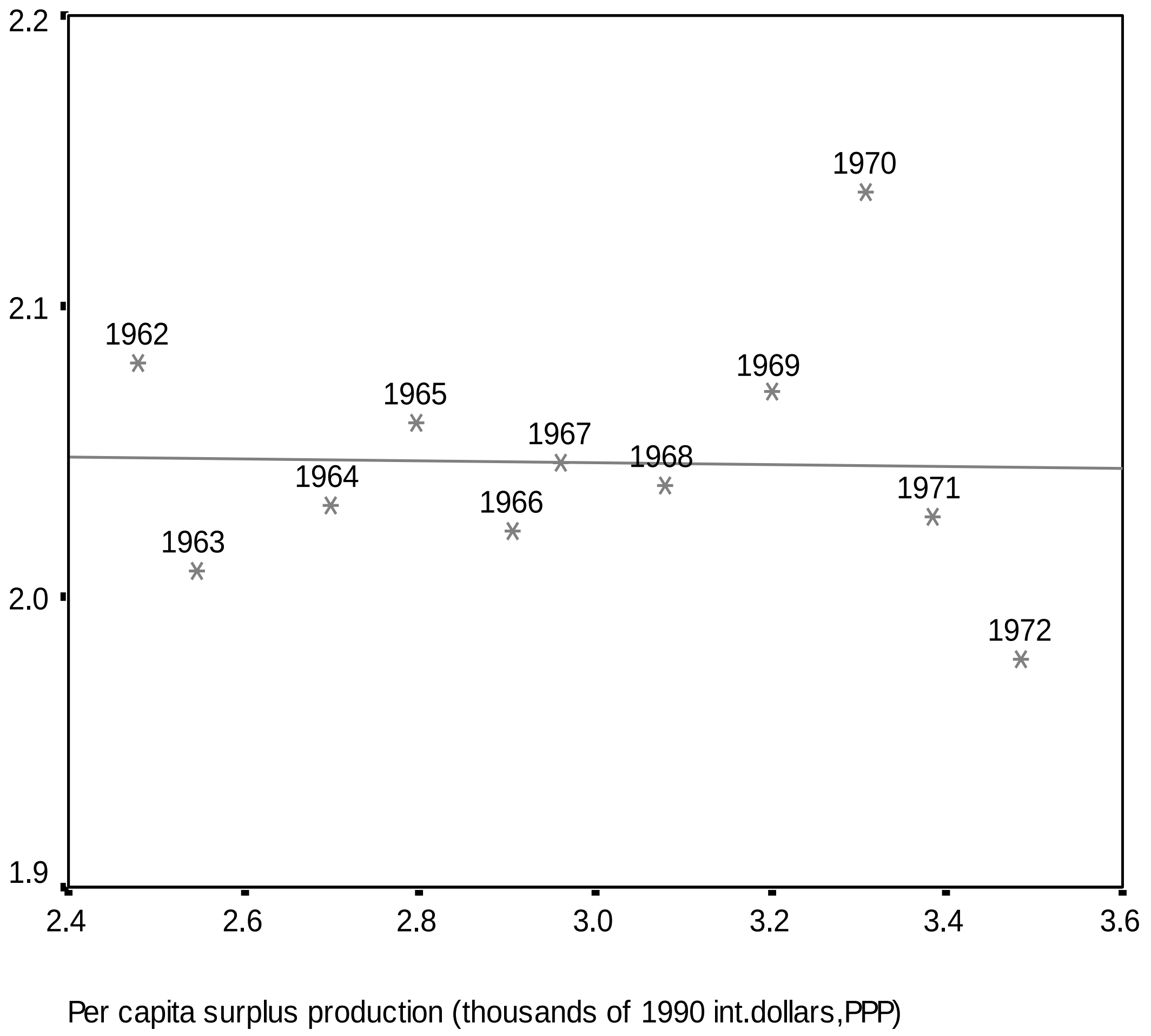

Fig. 12. Correlation between the per capita surplus production and world population growth rates for the range \$2400 < $S$ < \$3500 (scatterplot with fitted regression line). NOTES: $R = –.028$, $p = 0.936$. Data source – Maddison 2001.

Beyond $S = 3300$ ($g \sim 3700$) the correlation between $S$ and $r$ became strongly negative (see Fig. 13):

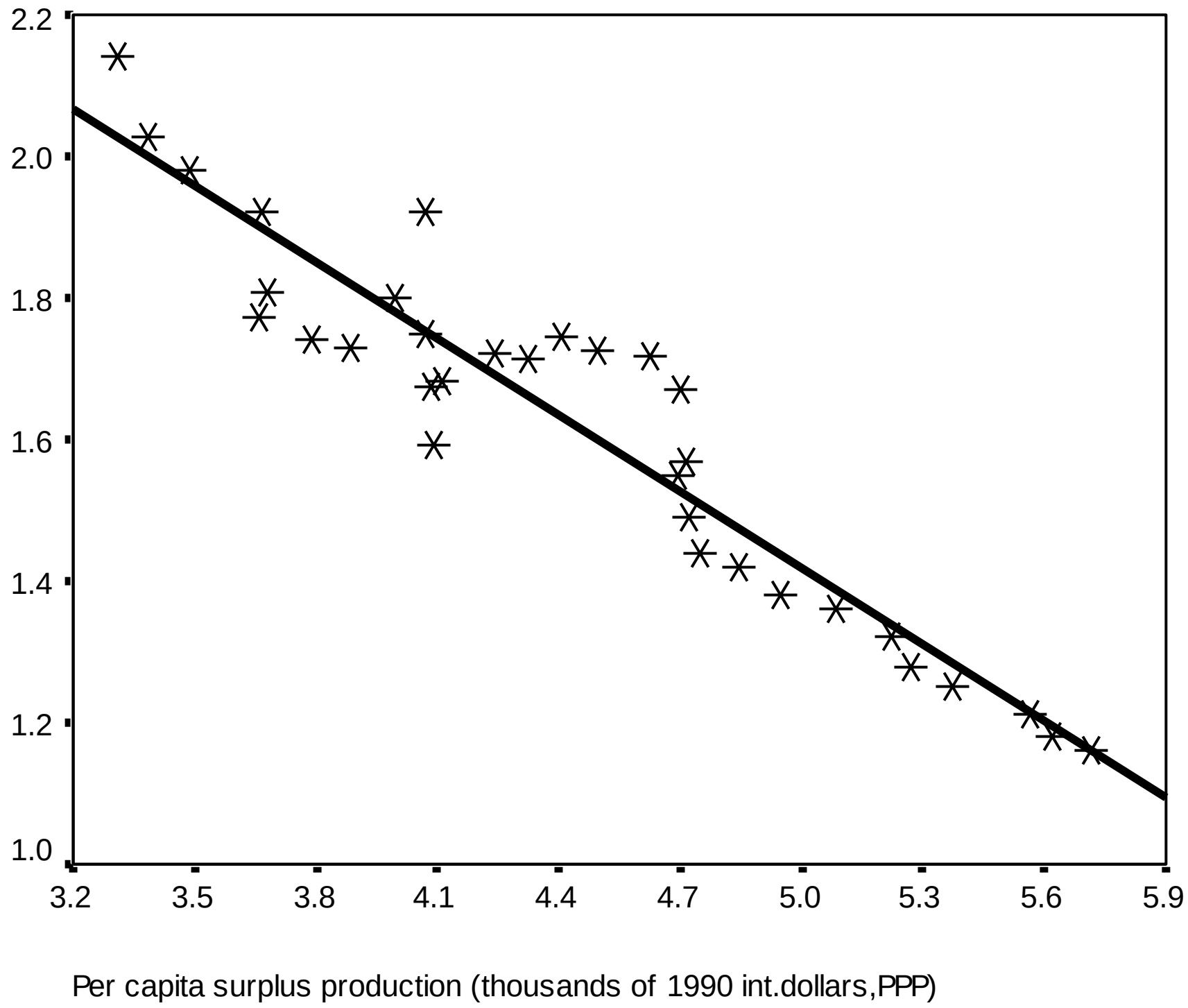

Fig. 13. Correlation between the per capita surplus production and world population growth rates for $S > \$4800$ (scatterplot with fitted regression line). NOTES: $R = –.946$, $p = .0000000000000003$. Data sources – [5] for the world per capita GDP, 1970–1998 and the world population growth rates, 1970–1988; [37] for the world per capita GDP, 1999–2002; [38] for the world population growth rates, 1989–2002.

What is more, in the range of $S > 4800$ ($g > 5200$) this negative correlation became almost perfect (see Fig. 14):

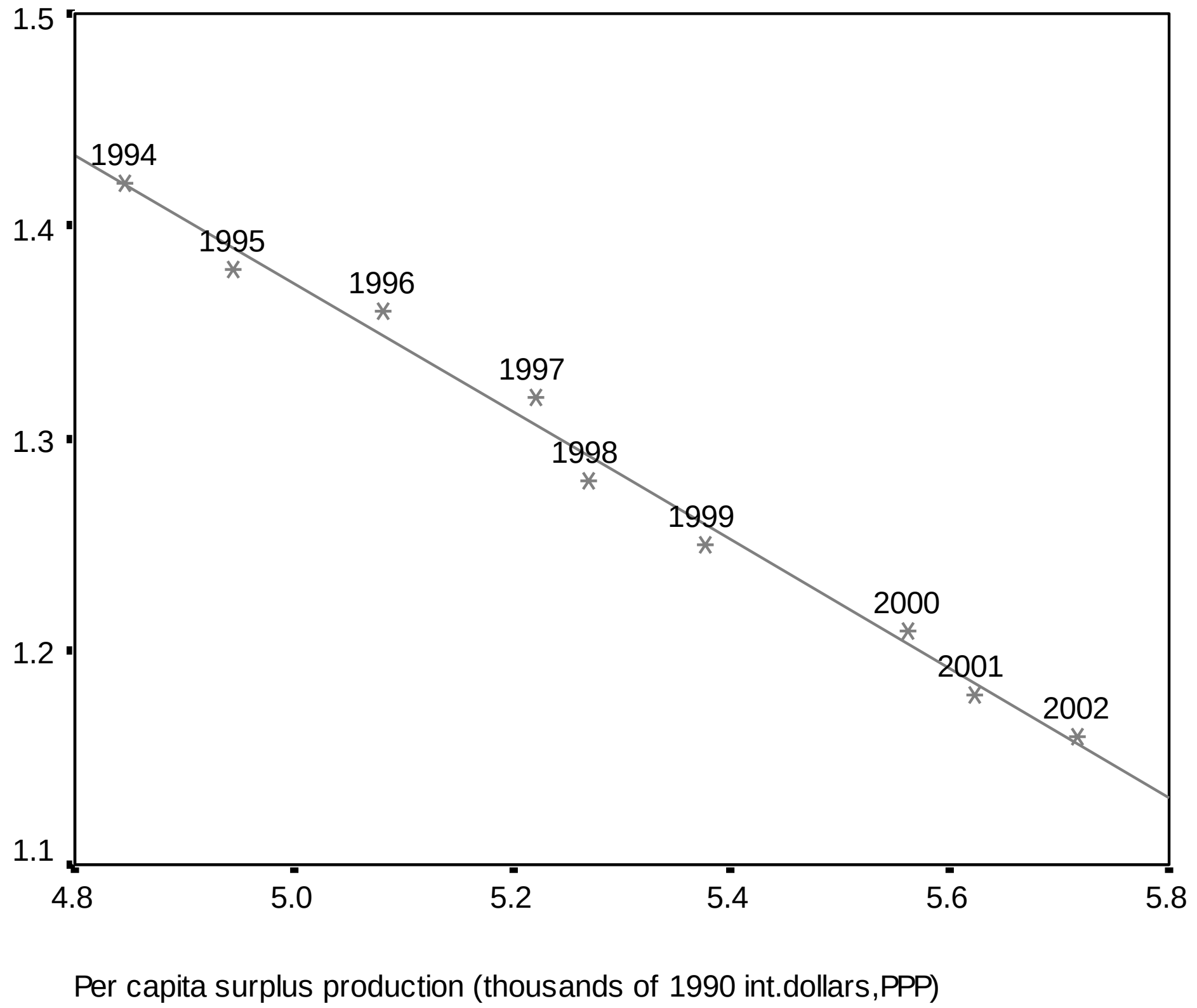

Fig. 14. Correlation between the per capita surplus production and world population growth rates for *S* > $4800 (scatterplot with fitted regression line). NOTES: *R* = –.995, *p* = .00000004. Data sources – [5] for the world per capita GDP, 1994–1998; [37] for the world per capita GDP, 1999–2002; [38] for the world population growth rates.

Thus, to describe the relationships between economic and demographic growth in the "post-Malthusian" GDP per capita range of > $3000, equation (17) should be modified, which is quite possible (see, e.g., [2,41]), but which goes out of the scope of this article aimed at the description of economic and demographic macrodynamics of the "Malthusian" period of the human history.

As was already noted by Kremer [7], in conjunction with equation (9) equation of (17) type "in the absence of technological change [that is if *T* = const] reduces to a purely Malthusian system, and produces behavior similar to the logistic curve biologists use to describe animal populations facing fixed resources" (actually, we would add to Kremer's biologists those social scientists who model pre-industrial economic-demographic cycles – see, *e.g.*, Refs. [42,43,44,45,46,47]).

Note that with a constant relative technological growth rate ( $\frac{\dot{T}}{T} = r_T = const$ ) within this model (combining eguations (9) and (17)) we will have both constant relative population growth rate ( $\frac{\dot{N}}{N} = r_N = const$ , and thus the population will grow exponentially) and constant *S*. Note also that the higher value of $r_T$ we take, the higher value of constant *S* we get.

Let us show this formally.

Take the following system:

$$G = gTN^{\alpha} \qquad (9)$$

$$\frac{dN}{dt} = aSN \qquad (17)$$

$$\frac{dT}{dt} = cT \ , \qquad (18')$$

where $S = \frac{G}{N} - m$ .

Equation (18') evidently gives

$$T = T_0 e^{ct} \ .$$

Thus,

$$G = gT_0 e^{ct} N^{\alpha}$$

and consequently

$$\frac{dN}{dt} = a\left(\frac{gT_0 e^{ct} N^{\alpha}}{N} - m\right)N = agT_0 e^{ct} N^{\alpha} - amN \ .$$

This equation is known as Bernoulli equation: $\frac{dy}{dx} = f(x)y + g(x)y^{\alpha}$ , which has the following solution:

$$y^{1-\alpha} = Ce^{F(x)} + (1-\alpha)e^{F(x)}\int e^{-F(x)}g(x)dx \ ,$$

where $F(x) = (1-\alpha)\int f(x)dx$ , and *C* is constant.

Is the case considered above, we have

$$N^{1-\alpha} = Ce^{F(t)} + (1-\alpha)e^{F(t)}\int e^{-F(t)}agT_0 e^{ct} dt \ ,$$

where $F(t)=(1-\alpha)\int(-am)dt=(\alpha-1)amt$.

So $N^{1-\alpha}=Ce^{(\alpha-1)amt}+(1-\alpha)agT_0e^{(\alpha-1)amt}\int e^{-(\alpha-1)amt}e^{ct}dt$

$$N^{1-\alpha}=e^{-(1-\alpha)amt}\left(C+(1-\alpha)agT_0\int e^{(c+(1-\alpha)am)t}dt\right)$$

$$N^{1-\alpha}=e^{-(1-\alpha)amt}\left(C+\frac{(1-\alpha)agT_0}{c+(1-\alpha)am}e^{(c+(1-\alpha)am)t}\right)$$

This result causes the following equation for *S*:

$$S=\frac{gT_0e^{ct}N^{\alpha}}{N}-m=gT_0e^{ct}N^{\alpha-1}-m=\frac{gT_0e^{(c+(1-\alpha)am)t}}{C+\frac{(1-\alpha)agT_0}{c+(1-\alpha)am}e^{(c+(1-\alpha)am)t}}-m$$

$$S=\frac{1}{\frac{C}{gT_0}e^{-(c+(1-\alpha)am)t}+\frac{(1-\alpha)a}{c+(1-\alpha)am}}-m$$

Since $c>0$ and $(1-\alpha)>0$, it is clear that $c+(1-\alpha)am>0$.

Consequently $e^{-(c+(1-\alpha)am)t}\to 0$ as $t\to\infty$.

This means that $S\xrightarrow[t\to\infty]{}\frac{c+(1-\alpha)am}{(1-\alpha)a}-m$, or finally

$$S\to\frac{c}{(1-\alpha)a} \text{ as } t\to\infty.$$

This, of course, suggests that in the growing "Malthusian" systems *S* could be regarded as a rather sensitive indicator of the speed of technological growth. Indeed, within Malthusian systems in the absence of technological growth the demographic growth will lead to *S* tending to 0, whereas a long-term systematic production of *S* will be only possible with systematic technological growth.[5]

[5] It might make sense to stress that it is not coincidental that we are speaking here just about the long-term perspective, as in a shorter-term perspective it would be necessary to take into account that within the actual Malthusian systems *S* was also produced quite regularly at the recovery phases of pre-Industrial political-demographic cycles (following political-demographic collapses as a result of which the surviving population found itself abundantly provided with resources). However, after the recovery phases the continuing production of significant amounts of *S* (and, hence, the continuing significant population growth) was only possible against the background of significant technological growth (see, *e.g.*, [41]). Note also that *S* produced at the initial (recovery) phases of political-demographic cycles in no way can explain the millennial trend towards the growth of *S* which was observed for many centuries before most of the world population moved to

Now replace $\frac{\dot{T}}{T} = r_T = const$ with Kremer's technological growth equation (13) and analyze the resultant model:

$$G = aTN^{\alpha}, \quad (9)$$

$$\frac{dN}{dt} = bSN, \quad (17)$$

$$\frac{dT}{dt} = cNT. \quad (13)$$

Within this model, quite predictably, $S$ can be approximated as $kr_T$. On the other hand, within this model, by definition, $r_T$ is directly proportional to $N$. Thus, the model generates an altogether not so self-evident (one could say even a bit unlikely) prediction – that throughout the "Malthusian-Kuznetsian" part of the human history the world per capita surplus production must have tended to be directly proportional to the world population size. This hypothesis, of course, deserves to be empirically tested. In fact, our tests have supported it.

Our test for the whole part of the human history, for which we have empirical estimates for both the world population and the world GDP (that is for 1–2002 CE)[6] has produced the following results: $R^2 = 0.98$, $p < 10^{-16}$, whereas for the period with the most pronounced "Malthusian-Kuznetsian" dynamics (1820–1958) the positive correlation between the two variables is almost perfect (see Fig. 15):

the second phase of demographic transition (e.g., [5]) and that appears to be produced just by the accelerating technological growth.

[6] Data sources – [5] for world population and GDP, 1–1998 CE, [37] for world GDP, 1999–2002, [38] for world population, 1999–2002.

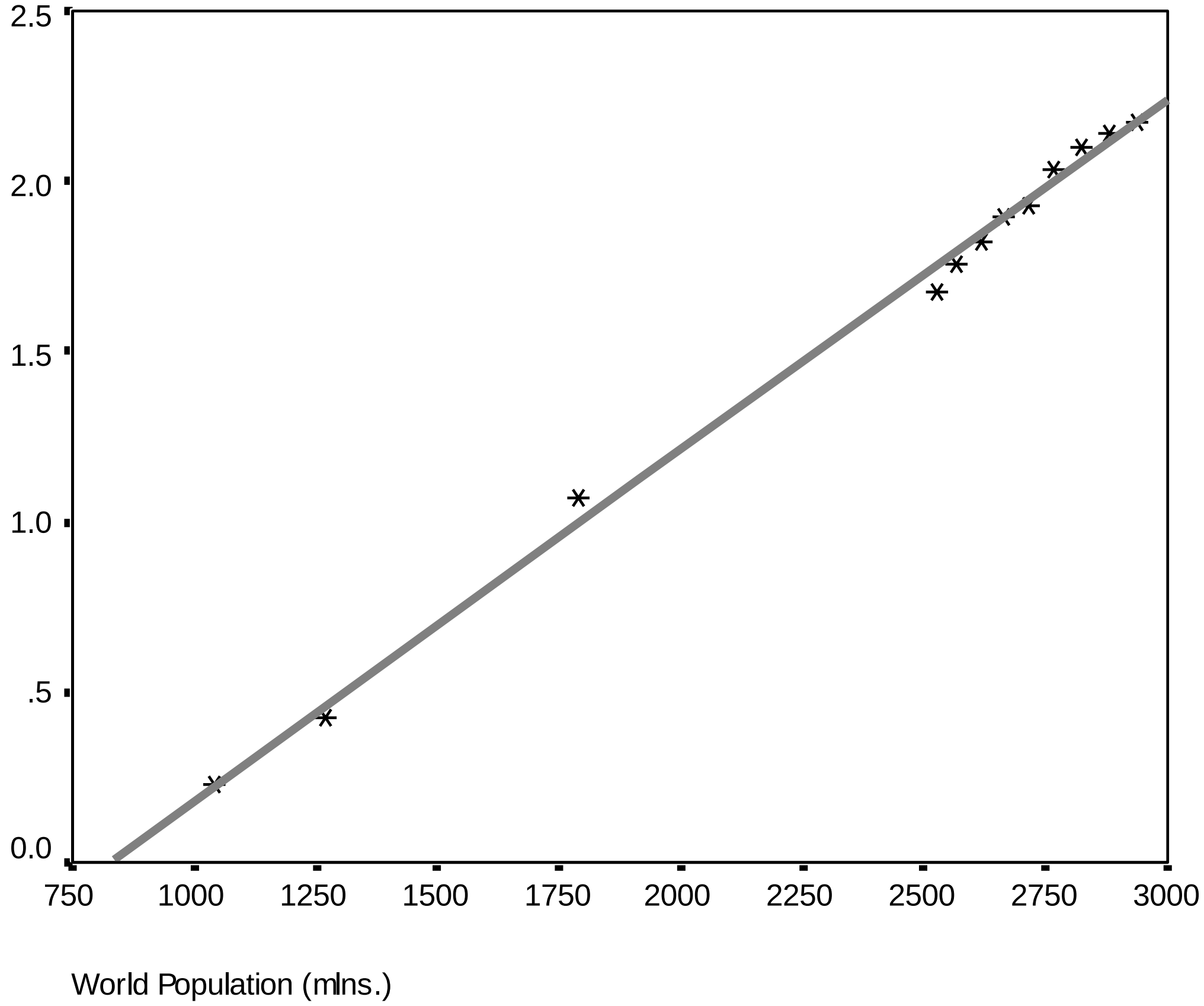

Fig. 15. Correlation between world population and per capita surplus production (1820–1958). NOTE: $R^2 > 0.996$, $p < 10^{-12}$.

Note that as within a Malthusian-Kuznetsian system $S$ can be approximated as $kN$, equation (17) may be appoximated as $dN/dt \sim k_1N^2$, or, of course, as $dN/dt \sim N^2 / C$; thus, Kapitza's equation turns out to be a by-product of the model under consideration.

Thus, we arrive at the following:

$$S \sim k_1 r_T \,,$$

$$r_T = k_2 N \,.$$

Hence,

$$dS/dt \sim kr_T/dt = k_3 dN/dt.$$

This implies that for the "Malthusian-Kuznetsian" part of human history $dS/dt$ can be approximates as $k_4dN/dt$.

On the other hand, as *dN/dt* in the original model equals *aSN*, this, of course, suggests that for the respective part of the human history both the economic and demographic World System dynamics may be approximated by the following unlikely simple mathematical model:[7]

$$\frac{dN}{dt} = aSN , \qquad (19)$$

$$\frac{dS}{dt} = bNS , \qquad (20)$$

where *N* is the world population, and *S* is surplus, which is produced per person with the given level of technology over the amount, which is minimally necessary to reproduce the population with a zero growth rate.

The world GDP is computed using the following equation:

$$G = mN + SN , \qquad (21)$$

where *m* denotes the amount of per capita GDP, which is minimally necessary to reproduce the population with a zero growth rate, and *S* denotes "surplus" produced per capita over *m* at the given level of the world-system techno-economic development.

Note that this model does not contain any variables, for which we do not have empirical data (at least for 1–1973) and, thus, a full empirical test for this model turns out to be perfectly possible.[8]

Incidentally, this model implies that the absolute rate of the world population growth (*dN/dt*) should have been roughly proportional to the absolute rate of the increase in the world per capita surplus production (*dS/dt*), and, thus (assuming the value of necessary product to be constant) to the absolute rate of the world per capita GDP growth, with which *dS/dt* will be measured thereafter. Note that among other things this could help us to determine the proportion between coefficients *a* and *b*.

Thus, if the model suggested by us has some correspondence to reality, one has grounds to expect that in the "Malthusian-Kuznetsian" period of the human history the absolute world population growth

[7] Note that this model only describes the Malthusian-Kuznetsian World System in a dynamically balanced state (when the observed world population is in a balanced correspondence with the observed technological level). To describe the situations with *N* disproportionally low or high for the given level of technology (and, hence, disproportionally high or low *S*) one would need, of course, the unapproximated version of the model ((9) – (17) – (13)). Note, that in such cases *N* will either grow, or decline up to the dynamic equilibrium level, after which the developmental trajectory will follow the line described by the (19) – (20) model.

[8] This refers particularly to the long-range data on the level of world technological development (*T*), which do not appear to be available till now.

rate (*dN/dt*) was directly proportional to the absolute growth rate of the world per capita surplus production (*dS/dt*). The correlation between these two variables looks as follows (see Fig. 16):

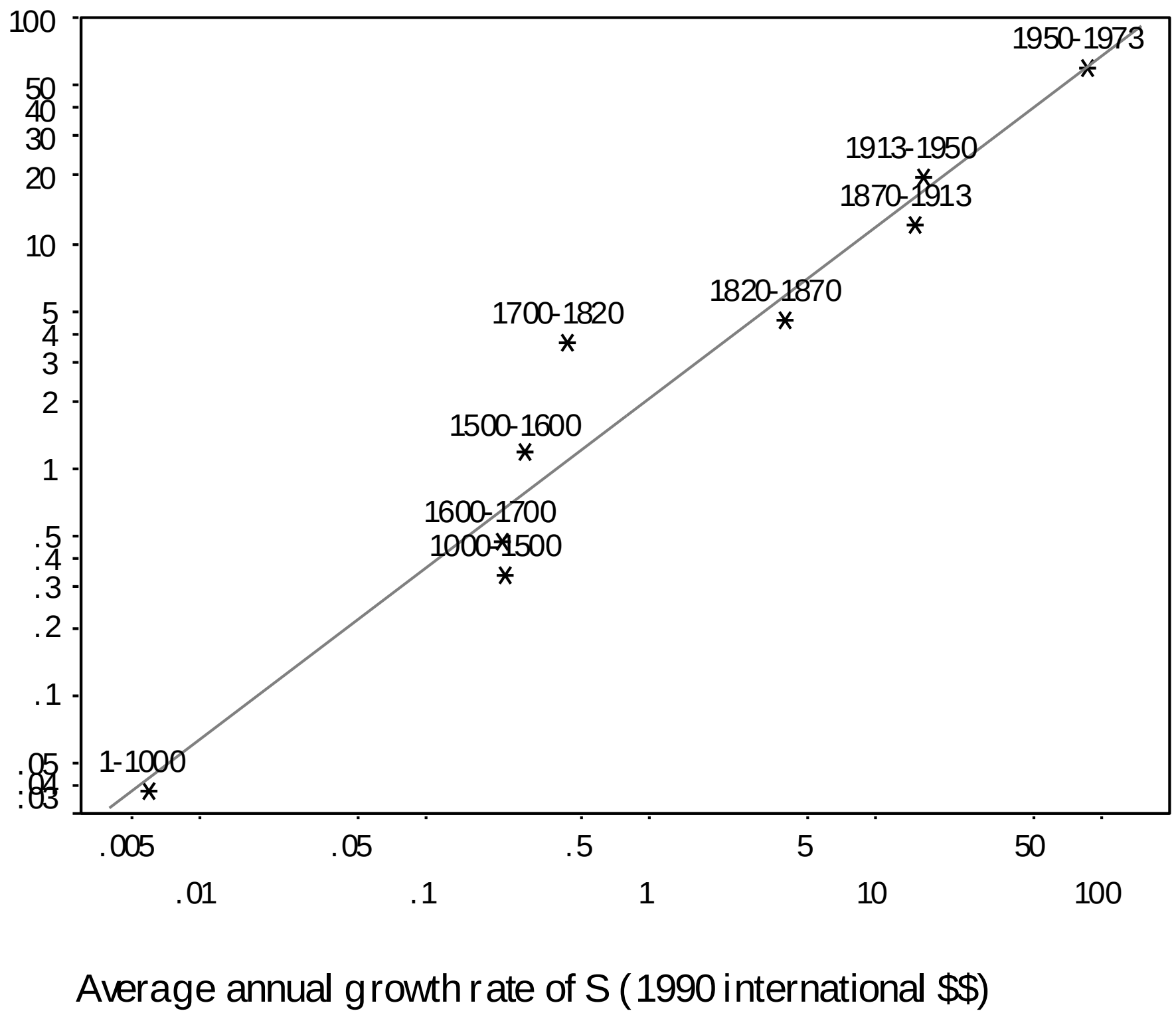

Fig. 16. Correlation between World Average Annual Absolute Growth Rate of Per Capita Surplus Production (*S*, 1990 PPP international dollars) and Average Annual Absolute World Population (*N*) Growth Rate (1 – 1973 CE), scatterplot in double logarithmic scale with a regression line. NOTE: Data source – [5]; Maddison's estimates of the world per capita GDP for 1000 CE has been corrected on the basis of Meliantsev [6].

Regression analysis of this dataset has given the following results (see Table 1):

Table 1.
Correlation between World Average Annual Absolute Growth Rate of Per Capita Surplus Production ($S$, 1990 PPP international dollars) and Average Annual Absolute World Population ($N$) Growth Rate (1 – 1950 CE), (regression analysis)

| | Unstandardized Coefficients | | Standardized Coefficients | t | Sig. |
|---|---|---|---|---|---|
| **Model** | **B** | **Std. Error** | **Beta** | | |
| **(Constant)** | 0.820 | 0.935 | | 0.876 | 0.414 |
| World Average Annual Absolute Growth Rate of Per Capita Surplus Production (1990 PPP international dollars a year) | 0.981 | 0.118 | 0.959 | 8.315 | <0.001 |
| **Dependent variable:** Average Annual Absolute World Population Growth Rate (mlns. a year) | | | | | |

NOTE: $R = 0.96$, $R^2 = 0.92$.

Note that the constant in this case is very small within the data scale, statistically insignificant, and lies within the standard error from 0, which makes it possible to equate it with 0. In this case regression analysis gives the following results (see Table 2):

Table 2.
Correlation between World Average Annual Absolute Growth Rate of Per Capita Surplus Production ($S$, 1990 PPP international dollars) and Average Annual Absolute World Population ($N$) Growth Rate (1 – 1950 CE), regression analysis, not including constant in equation

| | Unstandardized Coefficients | | Standardized Coefficients | t | Sig. |
|---|---|---|---|---|---|
| **Model** | **B** | **Std. Error** | **Beta** | | |
| World Average Annual Absolute Growth Rate of Per Capita Surplus Production (1990 PPP international dollars a year) | 1.04 | 0.095 | 0.972 | 10.94 | <0.001 |
| **Dependent variable:** Average Annual Absolute World Population Growth Rate (mlns. a year) | | | | | |

NOTE: $R = 0.97$, $R^2 = 0.945$.

Thus, just as implied by our second model, in the "Malthusian" period of the human history we do observe a strong linear relationship between the annual absolute world population growth rates ($dN/dt$) and the annual absolute growth rates of per capita surplus production ($dS/dt$). This relationship can be described mathematically with the following equation:

$$\frac{dN}{dt} = 1.04\frac{dS}{dt},$$

where $N$ is the world population (in millions), and $S$ is surplus (in 1990 PPP international dollars), which is produced per person with the given level of technology over the amount, which is minimally necessary to reproduce the population with a zero growth rate.

Note that according to model (19)-(20),

$$\frac{dN}{dt} = \frac{a}{b}\frac{dS}{dt}.$$

Thus, we get a possibility to express coefficient $b$ through coefficient $a$:

$$\frac{a}{b} = 1.04,$$

consequently:

$$b = \frac{a}{1.04} = 0.96a.$$

As a result, for the period under consideration it appears possible to simplify the second compact macromodel, leaving in it just one free coefficient:

$$\frac{dN}{dt} = aSN, \qquad (19)$$

$$\frac{dS}{dt} = 0.96aNS, \qquad (22)$$

With our two-equation model we start the simulation in the year 1 CE and do annual iterations with difference equations derived from the differential ones:

$$N_{i+1} = N_i + aS_iN_i,$$

$$S_{i+1} = S_i + 0.96aN_iS_i.$$

The world GDP is calculated using equation (21).

We choose the following values of the constants and initial conditions in accordance with historical estimates of Maddison (2001): $N_0$ = 230.82 (in millions); $a$ = 0.000011383; $S_0$ = 4.225 (in International 1990 PPP dollars).[9]

The outcome of the simulation, presented in Fig. 17 indicates that the compact macromodel in question is actually capable of replicating quite reasonably the world GDP estimates of Maddison (2001):

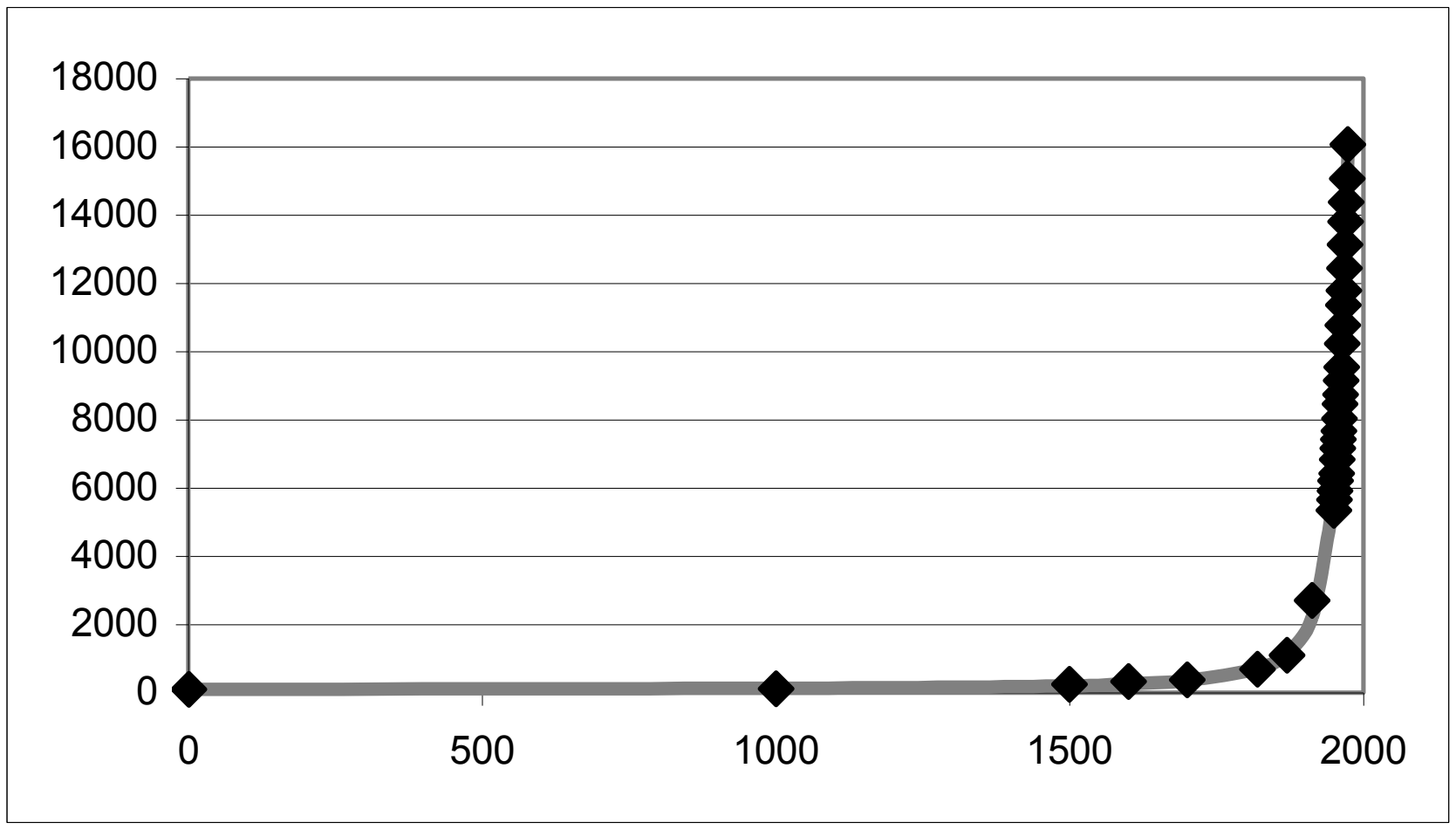

Fig. 17. Predicted and Observed Dynamics of the World GDP Growth, in billions of 1990 PPP international dollars (1 – 1973 CE). NOTE: The solid grey curve has been generated by the model; black markers correspond to the estimates of world GDP by Maddison [5].

The correlation between the predicted and observed values for this imitation looks as follows: $R > .999$; $R^2 = .9986$; $p << .0001$. For the world population these characteristics are also very high: $R$ = 0,996; $R^2$ = 0,992; $\alpha << 0,0001$.

According both to our model and the observed data up to the early 1970s we deal with the hyperbolic growth of not only the world population ($N$), but also per capita surplus production ($S$) (see Fig. 18):

[9] The value of $S_0$ was calculated with equation $S = G/N - m$ on the basis of Maddison's (2001) estimates for the year 1 CE. He estimates the world population in this year as 230.82 million, the world GDP as $102.536 billion (in 1990 PPP international dollars), and hence, the world per capita GDP production as $444.225 Maddison estimates the subsistence level per capita annual GDP production as $400 [5]. However, already by 1 CE most population of the world lived in rather complex societies, where the population reproduction even at zero level still required considerable production over subsistence level to maintain various infrastructures (transportation, legal, security, administrative and other subsystems etc.), without which even the simple reproduction of complex societies is impossible almost by definition. Note that the fall of per capita production in complex agrarian societies to subsistence level tended to lead to state breakdowns and demographic collapses (see, e.g., [42,43,44,46]). The per capita production to support the above mentioned infrastructures could hardly be lower than 10% of the subsistence level – that is close to Maddison's [5] of estimates, which makes it possible to estimate the value of $m$ as $440, and hence, the value of $S_0$ as $4.225.

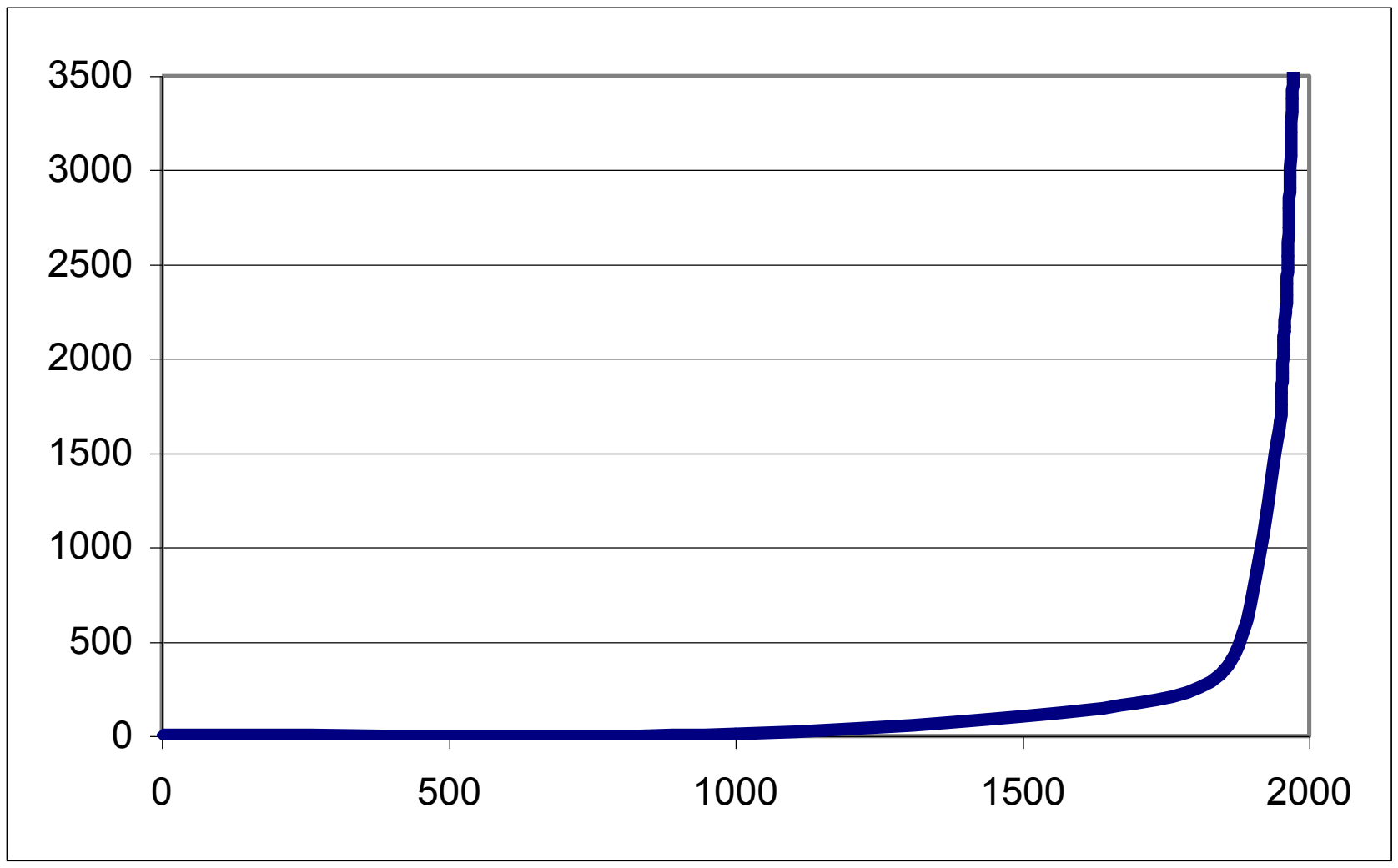

**Fig. 18.** Hyperbolic Growth of the World Per Capita Surplus Production, in 1990 PPP international dollars (1 – 1973 CE). NOTE: Data source – [5]; Maddison's estimates of the world per capita GDP for 1000 CE has been corrected on the basis of Meliantsev [6].

Note that even if *S* had not been growing, remaining constant, the world GDP would have been growing hyperbolically anyway through the hyperbolic growth of the world population only. However, the hyperbolic growth of *S* observed during this period of the human history led to the fact that the world population growth correlated with the world GDP growth not lineally, but quadratically (see Fig. 19):

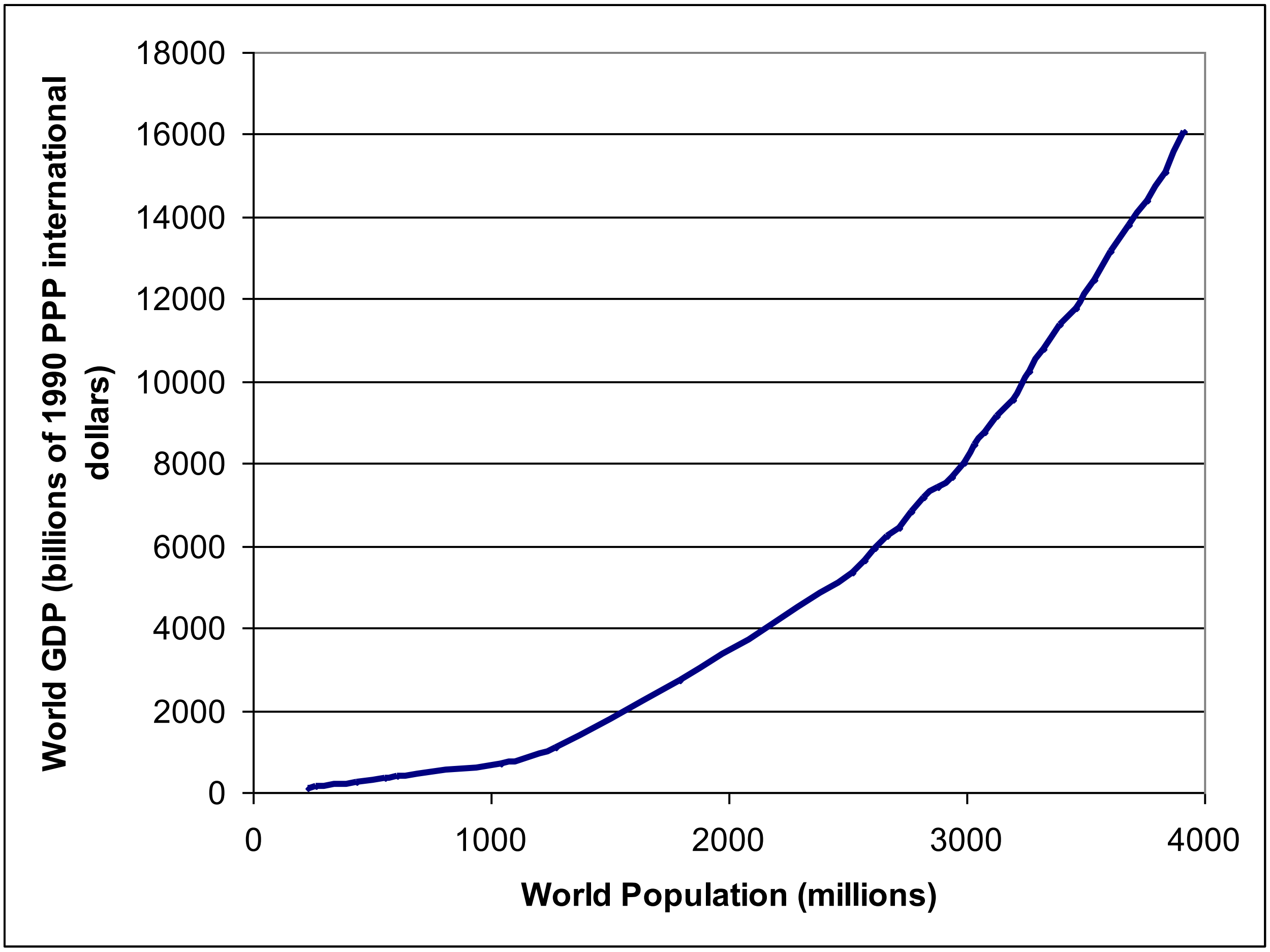

Fig. 19. Correlation between Dynamics of the World Population and GDP Growth (1 – 1973 CE). NOTE: Data source – [5]; Maddison's estimates of the world per capita GDP for 1000 CE has been corrected on the basis of Meliantsev [6].

Indeed, the regression analysis we have performed has shown here an almost perfect ($R^2$ = 0.998) fit just with the quadratic model (see Fig. 20):

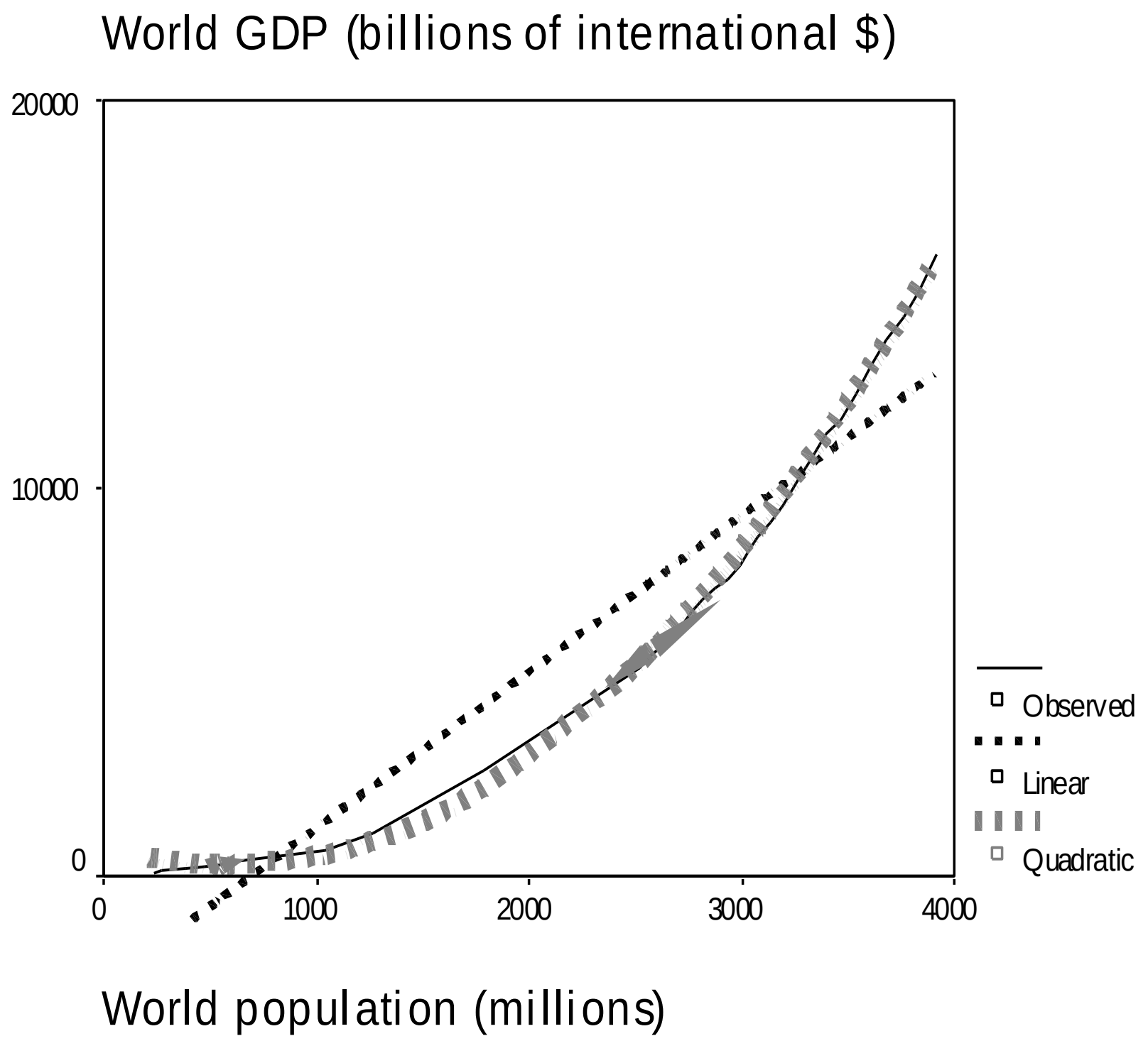

Fig. 20. Correlation between Dynamics of the World Population and GDP Growth (1 – 1973 CE): curve estimations

LINEAR REGRESSION: $\boldsymbol{R^2 = .876, p < .001}$

QUADRATIC REGRESSION: $\boldsymbol{R^2 = .998, p < .001}$

As a result the overall dynamics of the world GDP up to 1973 was not even hyperbolic, but rather quadratic-hyperbolic, leaving far behind the rather impressive hyperbolic dynamics of the world population growth (see Fig. 21):

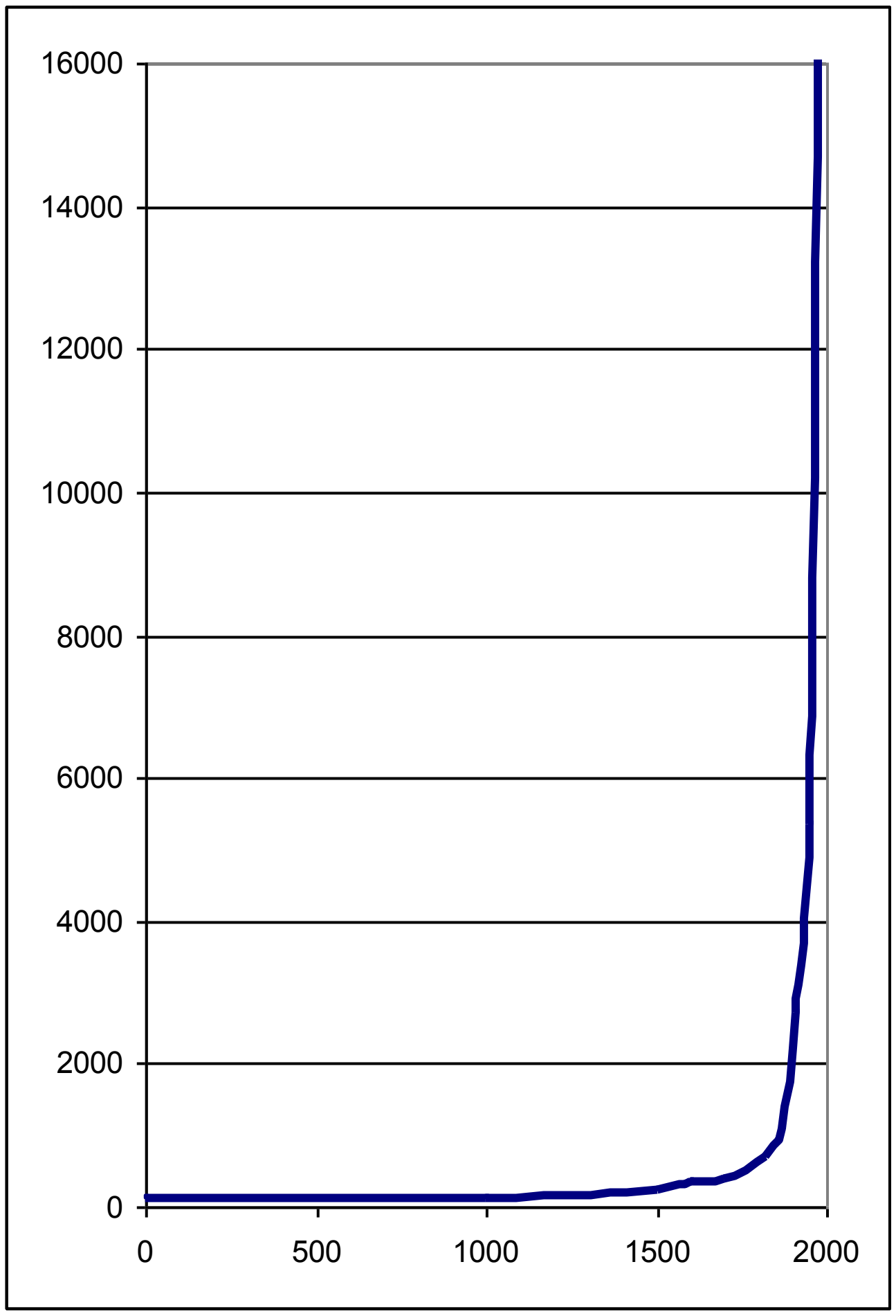

Fig. 21. The World GDP Growth from 1 CE up to the early 1970s (in billions of 1990 PPP international dollars) NOTE: Data source – [5]; Maddison's estimates of the world per capita GDP for 1000 CE has been corrected on the basis of Meliantsev [6].

The world GDP growth dynamics would look especially impressive if we take into account the quite plausible estimates of this variable change before 1 CE produced by DeLong [36], see Fig. 22:

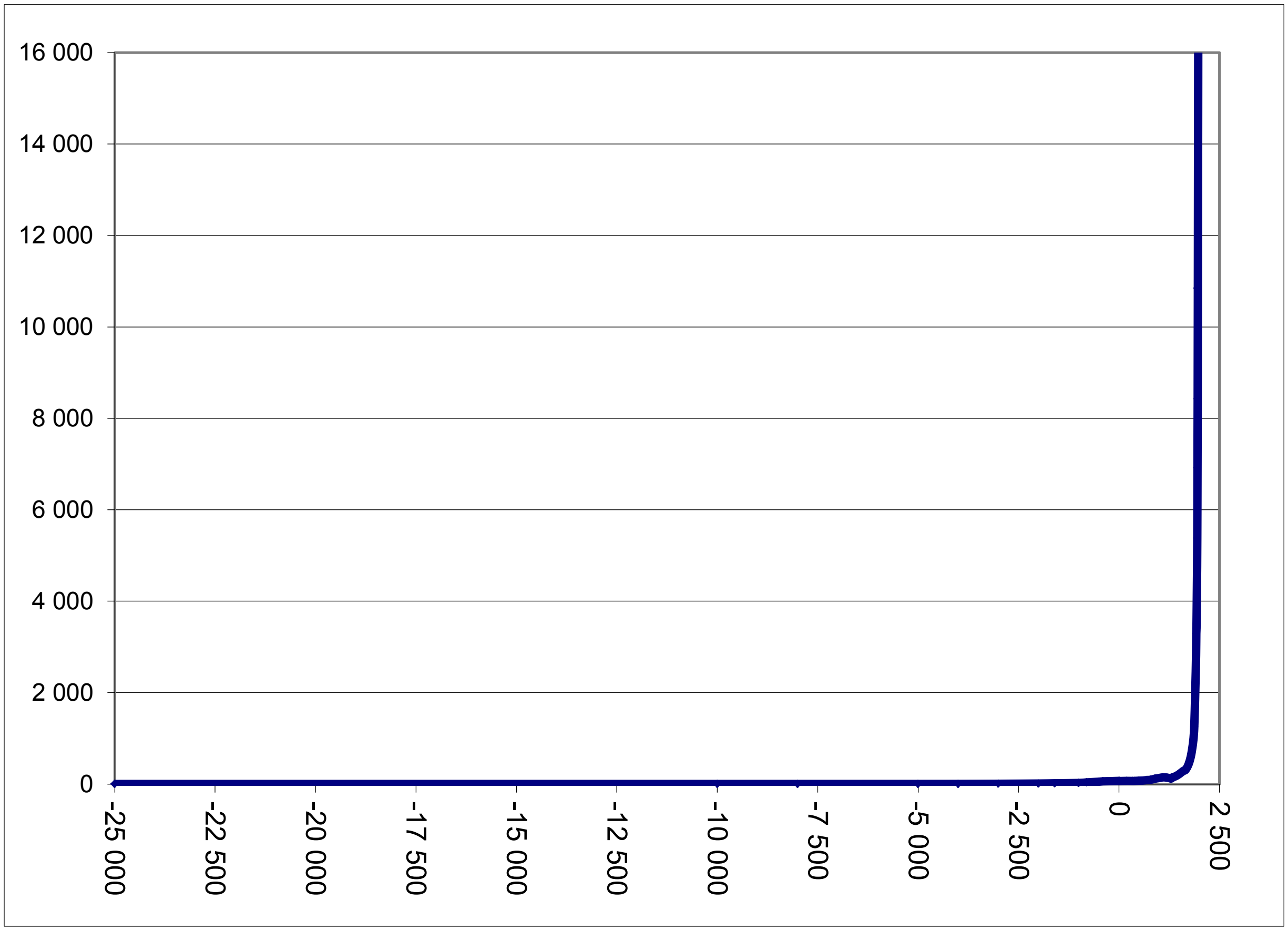

Fig. 22. The World GDP Growth from 25000 BCE up to the early 1970s (in billions of 1990 PPP international dollars)

It is fairly difficult not to admit that at this diagram the human economic history appears to be rather "dull" with the pre-Modern era looking as a period of almost complete economic stagnation, followed by the explosive modern economic growth. In reality the latter just does not let us discern at the diagram above the fact that many stretches of the pre-modern economic where characterized by dynamics that was comparatively no less dramatic. For example, as soon as we "zoom in" the apparently boringly flat stretch of the diagram above up to 800 BCE, we will see the following completely dynamic picture (see Fig. 23):

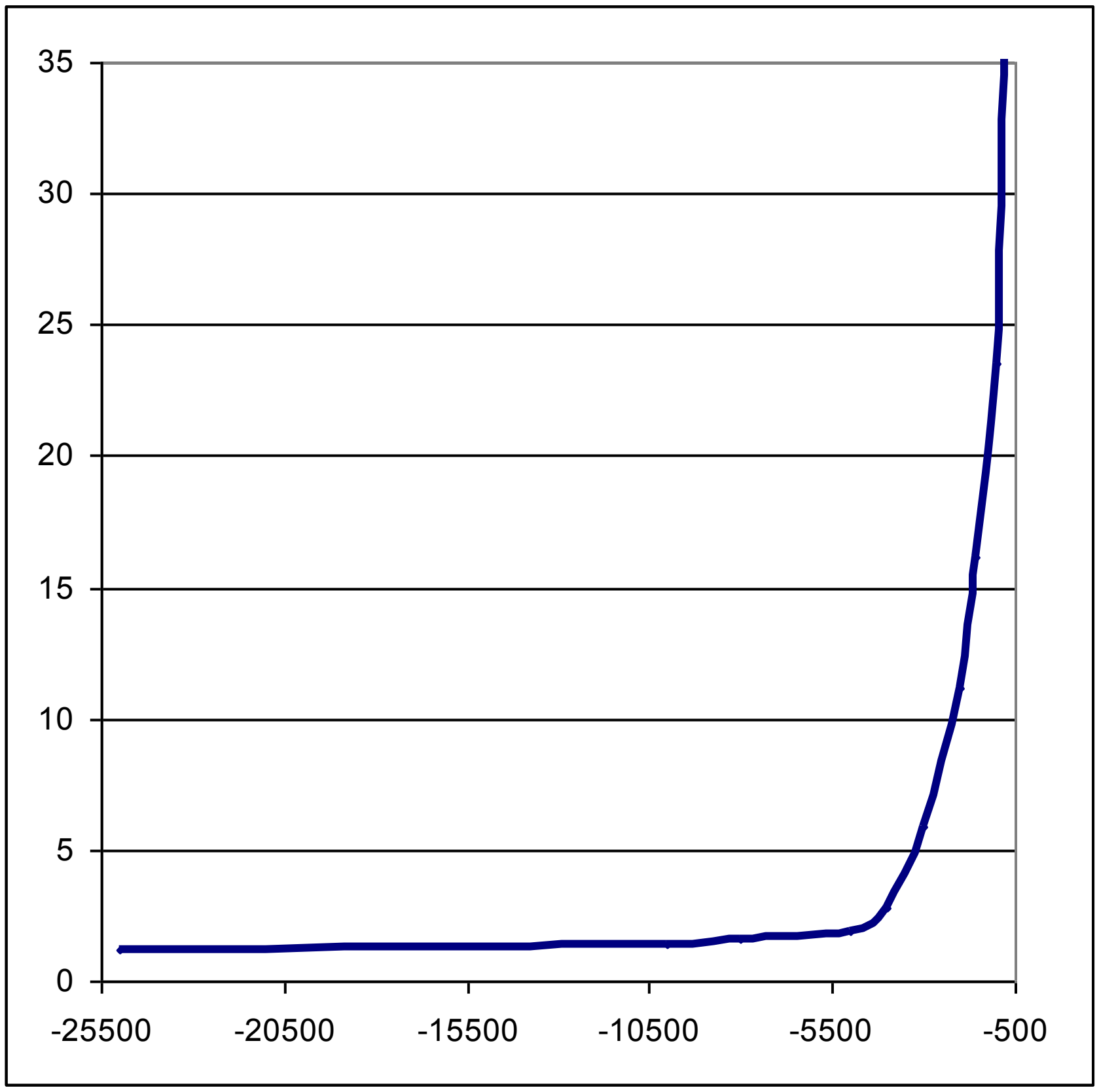

Fig. 23. The World GDP Growth from 25000 BCE up to 800 BCE (in billions of 1990 PPP international dollars)

This, of course, is accounted for by the difference in scales. During the "iron revolution" the World System economic growth was extremely fast in comparison with all the earlier epochs; however, according to DeLong's estimates in absolute scale it constituted less than $10 billion a century. At present the world GDP increases by $10 billion at average every three days [37]. As a result, even the period of relatively fast World System economic growth in the age of the "iron revolution" looks like almost horizontal line in comparison with the stretch of the modern economic growth. In other words the impression of the pre-modern economic stagnation created by Fig. 22 could be regarded as an illusion (in the strictest sense of this word) produced just by the quadratic-hyperbolic trend of the world GDP growth up to 1973.

Note also that we have already mentioned that as has been already shown by von Foerster [1], von Hoerner [3] and Kapitza [4] the world population growth before the 1970s is very well approximated by the following equation:

$$N = \frac{C}{t_0 - t}. \qquad (1)$$

As according to the model under consideration *S* can be approximated as *kN*, its long term dynamics can be approximated with the following equation:

$$S = \frac{kC}{t_0 - t}. \qquad (23)$$

Hence, the long-term dynamics of the most dynamical part of the world GDP, *SN*, the world surplus product, can be approximated as follows:

$$SN = \frac{kC^2}{(t_0 - t)^2}. \qquad (24)$$

As we could see at the beginning of this article, this approximation does work rather well.

## Conclusion

Human society is a complex nonequilibrium system that changes and develops constantly. Complexity, multivariability, and contradictoriness of social evolution lead researchers to a logical conclusion that any simplification, reduction, or neglect of the manifolds of factors leads inevitably to multiplication of error and to significantly erroneous understanding of processes under study. The view that any simple general laws are not observed at all with respect to social evolution has become totally predominant within the academic community, especially among those who specialize in the Humanities and who confront directly in their research all the manifolds and unpredictability of social processes. A way to approach human society as an extremely complex system is to recognize differences of abstraction and time scale between different levels. If the main task of scientific analysis is to detect the main acting forces so as to discover fundamental laws at a sufficiently coarse scale, abstracting from details and deviations from general rules at that level, then understanding at that level may help to identify measura-ble deviations from these laws in finer detail and faster time scales, not as a reductionism but contributing to measurement of deviations that are significant in their own right at finer and faster scales. Modern achievements in the field of mathematical modeling suggest that social evolution can be described with rigorous and sufficiently simple macrolaws.

As is well known in complexity studies – chaotic dynamics at the microlevel can generate a highly deterministic macrolevel behavior [45]. To describe behavior of a few gas molecules in a closed vessel we need very complex mathematical models, which will still be unable to predict long-run dynamics of such a system due to inevitable irreducible chaotic component. However, the behavior of zillions of gas molecules can be described with extremely simple sets of equations, which are capable of predicting almost perfectly the macrodynamics of all the basic parameters (and just because of chaotic behavior at microlevel). Of course, one cannot fail to wonder whether a similar set of regularities is not observed in the human world too, whether very simple regularities accounting for extremely high proportions of all the macrovariation cannot be found just for the largest possible social system – the World System.

Indeed, as we could see, the extremely simple mathematical models specified above can account for more than 99 per cent of all the variation in economic and demographic macrodynamics of the world for almost two millennia of its history.

In fact, this appears to suggest a novel approach to the formation of the general theory of social macroevolution. The approach prevalent in classical social evolutionism was based on an apparently self-evident assumption that evolutionary regularities of simple systems are significantly simpler than the ones characteristic for complex systems. A rather logical outcome from this almost self-evident assumption is that one should study first evolutionary regularities of simple systems and only after understanding them to move to more complex ones.[10] One wonders if the opposite direction might not be more productive – from the study of simple laws of the development of the most complex social system to the study of the complex regularities of evolution of simple social systems.

In general, it can be demonstrated [2,34,35,44] that the hyperbolic pattern of the world's population growth could be accounted for by the nonlinear second order positive feedback mechanism that was shown long ago to generate just the hyperbolic growth, known also as the ‘blow-up regime’ [48]. In our case this nonlinear second order positive feedback looks as follows: the more people – the more poten-

[10] Of course, a major exception here is constituted by the world-system approach (*e.g.*, [31,32,33]), but the research of world-system students has by now yielded somehow limited results, to a significant extent because they have not used sufficiently standard scientific methods implying that verbal constructions should be converted into mathematical models, whose predictions are to be tested with available data.

tial inventors – the faster technological growth – the faster growth of the Earth's carrying capacity – the faster population growth – with more people you also have more potential inventors – hence, faster technological growth, and so on (see Fig. 24):

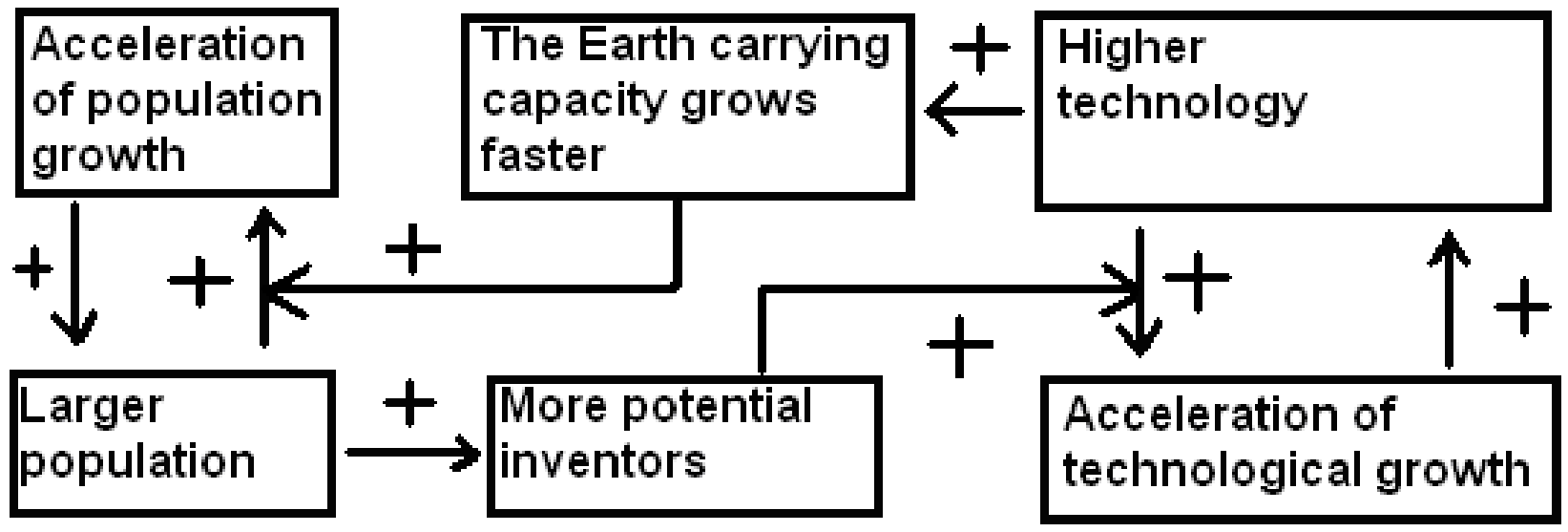

Fig. 24. Block scheme of the nonlinear second order positive feedback between technological development and demographic growth

Note that the relationship between technological development and demographic growth cannot be analyzed through any simple cause-and-effect model, as we observe a true dynamic relationship between these two processes – each of them is both the cause and the effect of the other. Up to the 1970s the hyperbolic growth of the world population was accompanied by the quadratic-hyperbolic growth of the world GDP, just as is suggested by our model. Note that the hyperbolic growth of the world population and the quadratic-hyperbolic growth of the world GDP are also very tightly connected processes, actually two sides of the same coin, two dimensions of one process propelled by the nonlinear second order positive feedback loops between the technological development and demographic growth (see Fig. 25):

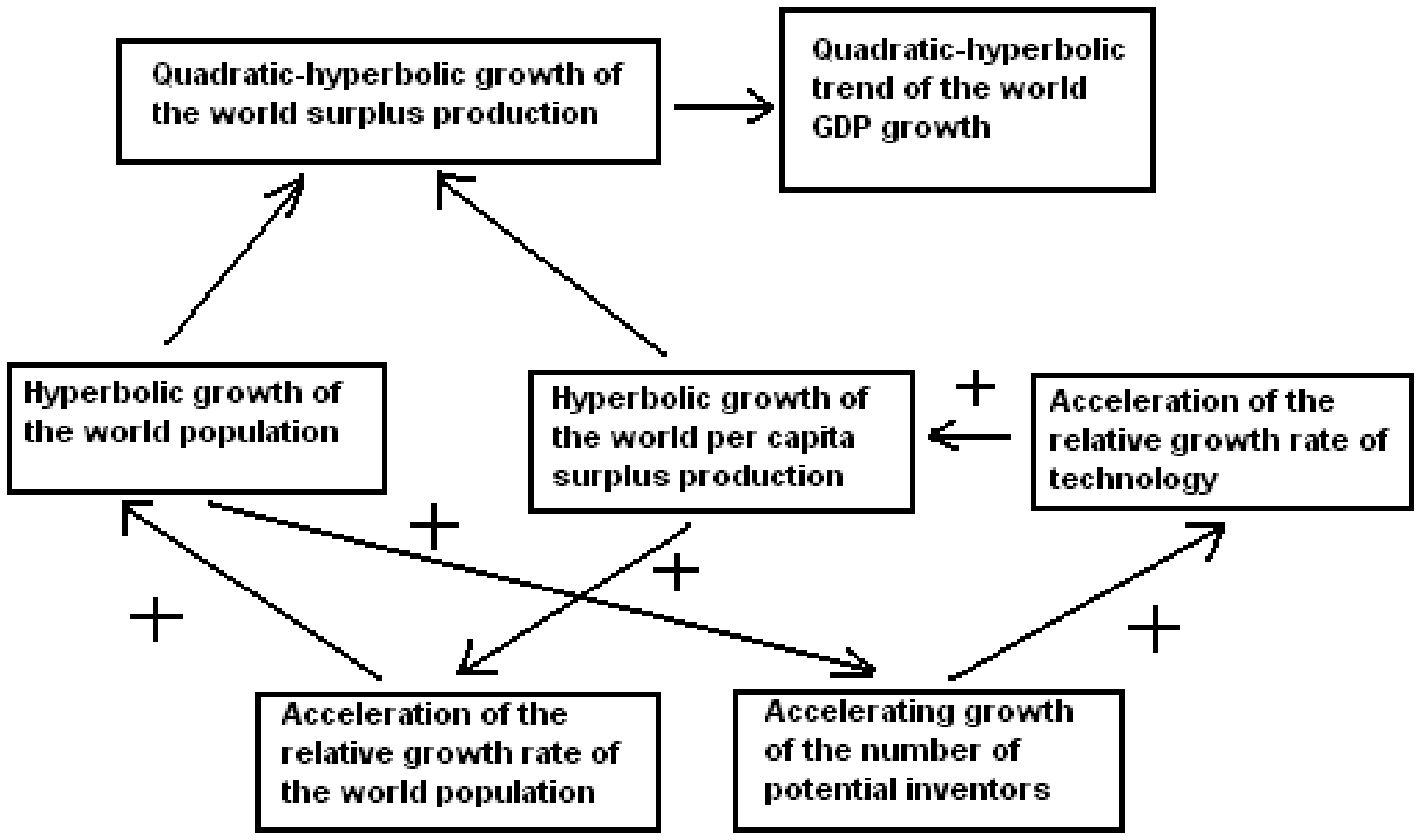

Fig. 25. Block Scheme of the Generation of Quadratic-Hyperbolic Trend of the World Economic Growth by the Nonlinear Second Order Positive Feedback between Technological Development and Demographic Growth